\documentclass{article}
\usepackage[english]{babel}

\usepackage{amsmath}
\usepackage{amssymb}

\usepackage{graphicx}
\usepackage[colorlinks=true,linkcolor=cyan, citecolor=cyan, urlcolor=magenta]{hyperref}      

\usepackage{caption}
\usepackage{subcaption}
\usepackage{chngcntr}
\usepackage{caption}

\begin{document}

\title{{\Large Dividing goods or bads under additive utilities\thanks{%
Support from the Basic Research Program of the National Research University
Higher School of Economics is gratefully acknowledged. Moulin thanks the
Simmons institute for Theoretical Computing for its hospitality during the
Fall 2015. Sandomirskiy is partially supported by the grant 16-01-00269 of
the Russian Foundation for Basic Research. The comments of William Thomson
on an earlier version have been especially useful.}}}
\author{{\large Anna Bogomolnaia}$^{\bigstar \spadesuit }${\large , Herv\'{e}
Moulin}$^{\bigstar \spadesuit }${\large ,} \and {\large Fedor Sandomirskiy}$%
^{\spadesuit } ${\large , and Elena Yanovskaia}$^{\spadesuit }$}
\date{{\large $^{\bigstar } $ \textit{University of Glasgow}\\
\vskip 0.1cm 
$^{\spadesuit }$ 
\textit{Higher School of Economics, St Petersburg}}\\
\vskip 0.9cm
Revised, {\large May 2017}}
\maketitle

\begin{abstract}
We compare the \textit{Egalitarian Equivalent }and the \textit{Competitive
Equilibrium with Equal Incomes }rules to divide a bundle of goods
(heirlooms) or a bundle of bads (chores).

For goods the Competitive division fares better, as it is Resource
Monotonic, and makes it harder to strategically misreport preferences. But
for bads, the Competitive rule, unlike the Egalitarian one, is multivalued,
harder to compute, and admits no continuous selection.

We also provide an axiomatic characterization of the Competitive rule based
on the simple formulation of Maskin Monotonicity under additive utilities.
\end{abstract}

\section{Introduction}

User-friendly platforms like SPLIDDIT, Adjusted Winner, or The Fair Division
Calculator\footnote{%
\url{www.spliddit.org/}; \url{www.nyu.edu/projects/adjustedwinner/}; \url{www.math.hmc.edu/~su/fairdivision/calc/}} implement theoretical solutions to a
variety of fair division problems, among them the classic distribution of a
bundle of divisible commodities (the \textquotedblleft
manna\textquotedblright ). The key simplification is that these platforms
ask visitors to report linear preferences (additive utilities), instead of
potentially complex Arrow-Debreu preferences. Say we divide the family
heirlooms: each participant on SPLIDDIT must distribute 1000 points over the
different objects, and these \textquotedblleft bids\textquotedblright\ are
interpreted as her \textit{fixed} marginal rates of substitution\textit{. }%
Eliciting complementarities between these objects is potentially a complex
task with 6 objects, an outright impossible one with 10 or more, hence the
design choice of deliberately ignoring them. For the same reason
combinatorial auction mechanisms never ask buyers to report a ranking of all
subsets of objects, (\cite{BL}, \cite{VV}, \cite{CSS}). The proof of the
pudding is in the eating: \ visitors use these sites in the tens of
thousands, fully aware of the interpretation of their bids (\cite{GP}).

The two theoretical division rules used on the first two sites above are the 
\textit{Competitive Equilibrium with Equal Incomes} (for short, \textit{%
Competitive})\textit{\ }rule and the \textit{Egalitarian Equivalent} (for
short \textit{Egalitarian}) rule: see \cite{V} and \cite{PS} respectively.
The latter finds an efficient allocation where everyone is indifferent
between his share and a common fraction of the entire manna. The former
identifies prices of the commodities and a common budget constraint at which
the competitive demands are feasible, and implement these demands. Here we
critically compare the performance of these two rules in the additive domain.

Fair division problems may involve \textit{bads} (non disposable items
generating disutility) as well as goods (disposable, desirably commodities).
For the former think of workers distributing tasks (house chores, job shifts
among substitutable workers (\cite{B}) like teaching loads, babysitting,
etc.), cities sharing noxious facilities, managers allocating cuts in the
company's workforce between their respective units, and so on. For the
latter, the family heirlooms (\cite{PZ}), the assets of divorcing partners (%
\cite{BT}), office space between the colleagues, seats in overdemanded
business school courses (\cite{SU}, \cite{BC}), computing resources in
peer-to-peer platforms (\cite{GZHKSS}), and so on.

If we divide \textit{goods} both rules, Competitive and Egalitarian, are
single-valued (utilitywise), easy to compute, and vary continuously in the
marginal utility parameters. This is clear for the latter rule, and for the
former one it follows from the celebrated Eisenberg-Gale theorem (\cite{EG}, 
\cite{SS}): the Competitive allocations also maximize the Nash product of
utilities over all feasible allocations. We invoke two normative properties
to argue that the Competitive rule outperforms the Egalitarian rule. One is
the familiar \textit{Resource Monotonicity }(\textbf{RM}), stating that an
increase of the \textquotedblleft good\textquotedblright\ manna should
improve at least weakly everyone's welfare. The Competitive rule meets RM
but the Egalitarian does not.

The second axiom is new, though it can be traced to Maskin Monotonicity (see
Subsection 5.3). \textit{Independence of Lost Bids} (\textbf{ILB}) is
predicated on the observation that, when utilities are additive, at an
efficient allocation most of the entries in the consumption matrix
(specifying how much each agent eats of each good) are zero. If agent $i$
does not consume good $a$ at all, we call his marginal utility $u_{ia}$ a
\textquotedblleft lost bid\textquotedblright . ILB states that changing $i$%
's lost bid $u_{ia}$ should have no effect on the outcome, as long this bid
remains lost. This property has important incentives consequences. Recall
that, even in the additive domain,no reasonable division rule (e. g.,
treating equals equally) can be both efficient and strategyproof (\cite{Sch}%
). A consequence of ILB is that an agent cannot benefit by a \textit{small}
misreport of her lost bids. The Competitive rule meets ILB: it is still
vulnerable to misreports of one's winning bids, but the profitable direction
of misreports depends upon the entire problem, so it requires a fair amount
of information. By contrast, the Egalitarian rule fails ILB, and the (small)
profitable misreports are entirely clear: one should inflate lost bids and
deflate winning bids: Lemma 4 in Subsection 5.3.

Combined with efficiency and symmetry properties, ILB characterizes the
Competitive division rule.

Turning to the division of \textit{bads} we find, somewhat surprisingly,
that the Competitive rule is much harder to handle. The main difficulty is
that we may have many different competitive utility profiles, even
exponentially many in (the smallest of) the number of bads and of agents
(Subsection 5.4). There is no obvious way to deal with this embarrassing
multiplicity. In particular every selection of the Competitive
correspondence is discontinuous in the parameters (marginal utilities) of
the economy, and this is also true of any selection of the much larger
correspondence of efficient and Envy Free allocations. The (long) proof uses
the fact that the set of efficient and envy-free allocations, and the
corresponding disutility profiles, can have close to $\frac{2}{3}n$
connected components. Finally, computing the competitive allocations of bads
is not a convex optimization problem as in the case of goods, and with more
than two agents we do not know of any efficient algorithms discovering them.

By contrast the Egalitarian rule to divide bads is the mirror image of the
rule for goods, and shares the same properties: it is single-valued and
continuous in the utility parameters (as well as in the manna). We conclude
that the Egalitarian is a more practical approach to the division of bads
than the Competitive one, or any Envy-Free single-valued rule.

\paragraph{Contents}

After the literature review in Section 2, the model is defined in Section 3,
and our two division rules in Section 4. Section 5 contains our normative
comparison of the two rules. The Egalitarian rule ensures to each agent
strictly more than her \textit{Fair Share }utility or disutility (Subsection
5.1). The Competitive rule for goods is Resource Monotonic (Subsection 5.2),
and satisfies Independence of Lost Bids (subsection 5.3). We divide bads in
the last two Subsections: we discuss successively the multiplicity issue
(5.4) then the discontinuity issue (5.5). All substantial proofs are in
Section 6\textbf{.}

\section{Related literature}

\hskip 0.5cm 1. Our main motivation is the recent stream of work in algorithmic mechanism
design on the fair division of \textit{goods}, recognizing the practical
convenience of additive utilities and the conceptual advantages of the
Competitive solution. For instance in the same model as here, Megiddo and
Vazirani (\cite{MV}) show that the Competitive utility profile depends
continuously upon the rates of substitution and the total endowment; Jain
and Vazirani (\cite{JV}) that it is can be computed in time polynomial in
the dimension $n+p$ of the problem.

Steinhaus' 1948 \textquotedblleft cake-division\textquotedblright\ model (%
\cite{St}), also assumes linear preferences represented by atomless measures
over, typically, a compact euclidean set. It contains our model for goods as
the special case where the measures are piecewise constant. Sziklai and
Segal-Halevi (\cite{SS1}) show that the Eisenberg-Gale Theorem still holds,
and that the Competitive rule is Resource Monotonic (see the Remark in
Subsection 6.3).\smallskip

2. In the companion paper \cite{BMSY} we consider the more general problem
of dividing a \textquotedblleft mixed manna\textquotedblright\ containing
both goods and bads, as when we dissolve a partnership with both valuable
assets and liabilities. Our first observation is that the Egalitarian rule
is no longer well defined, because there may be no efficient allocation
where everybody is indifferent to consuming a common fraction of the entire
manna (or of any common benchmark bundle). So the competitive rule wins our
contest by default.

The main message of \cite{BMSY} is that mixed manna problems are of two
types. If goods overwhelm bads\footnote{%
In the sense that some feasible division of the manna gives everyone a
positive utility.} the Competitive rule behaves just like an all goods
problem: it picks a maximizes the product of utilities, yields a unique
utility profile, is resource monotonic and continuous. But if instead bads
overwhelm goods we are back to the potentially messy situation of an all
bads problems with a host of different competitive divisions and no
continuous selection from this set.\smallskip

3. Four decades earlier, the microeconomic literature on the fair division
of private goods insisted on working in the much larger domain of
Arrow-Debreu preferences, where the relation between the Nash product of
utilities and the Competitive rule is lost, and provided several axiomatic
characterizations of the latter. The most popular result appears first in
Hurwicz\textbf{\ }(\cite{H}) and Gevers (\cite{G}), and is refined by Thomson%
\textbf{\ }(\cite{T2}) and Nagahisa (\cite{N}): any efficient and Pareto
indifferent rule meeting (some variants of) Maskin Monotonicity (MM) must
contain the Competitive rule. Our Independence of Lost Bids axiom is in fact
a weak variant of MM for the linear domain, and the proof of our
characterization result (Proposition 3) mimics the standard argument.

\section{Division problems and division rules}

The finite set of agents is $N$ with generic element $i$, and $|N|=n\geq 2$.
The finite set of divisible items is $A$ with generic element $a$ and $%
|A|=p\geq 2$. The manna consists of one unit of each item. We assume the
manna contains either only goods, or only bads, and we use the same notation
for both types of problems.

Agent $i$'s allocation (or share) is $z_{i}\in \lbrack 0,1]^{A}$; the
profile $z=(z_{i})_{i\in N}$ is a feasible allocation if $%
\sum_{N}z_{i}=e^{A} $, the vector in $%
\mathbb{R}
_{+}^{A}$ with all coordinates equal to $1$. The set of feasible allocations
is $\Phi (N,A)$.

Each agent is endowed with linear preferences over $[0,1]^{A}$, represented
by a vector $u_{i}\in 
\mathbb{R}
_{+}^{A}$, a utility function in the case of goods, a disutility function in
that of bads. We use the generic term \textit{utility}$^{\ast }$ for both
cases, which will generate no confusion. Only the underlying\textit{\ }%
preferences matter: for any $\lambda >0$, $u_{i}$ and $\lambda u_{i}$ carry
the same information. This restriction is formally included in Definition 1
below.

Given an allocation $z$ we write agent $i$'s corresponding \textit{utility}$%
^{\ast }$ as $U_{i}=u_{i}\cdot z_{i}=\sum_{A}u_{ia}z_{ia}$.

Clearly a \textquotedblleft null agent\textquotedblright\ ($\forall
a:u_{ia}=0$) can be ignored when we divide goods. When we divide bads, the
problem is trivial if there is a null agent ($U=0$ is feasible and uniquely
efficient). Thus we only look at problems where all agents are non null.
Moreover if we divide bads, the vector $U=0$ is feasible if (and only if)
each bad is harmless to at least one agent ($\forall a\exists i:u_{ia}=0$):
such problems are also trivial and we also rule them out in the Definitions
below.

Similarly if item $a$ gives $u_{ia}=0$ for all $i$, it is a
\textquotedblleft useless good\textquotedblright\ or a \textquotedblleft
harmless bad\textquotedblright\ that can be ignored as well. Our Competitive
rule to divide bads goes one step further: it will also ignore a bad $a$
harmless to \textit{some} agents, and give no credit to these agents for
eating $a$. See Definition 4.\smallskip

\noindent \textbf{Definition 1}

\noindent \textit{A division problem is a triple }$\mathcal{Q}=(N,A,u)$%
\textit{\ where }$u\in 
\mathbb{R}
_{+}^{N\times A}$\textit{\ is such that the }$N\times A$\textit{\ matrix }$%
[u_{ia}]$\textit{\ has no null row, no null column, and in the case of bads
there is at least one column with no null entry.}

\noindent \textit{We write }$\Psi (\mathcal{Q})$ \textit{for the set of
feasible \textit{utility}$^{\ast }$ profiles, and} $\Psi ^{eff}(\mathcal{Q})$%
\textit{for its subset of efficient \textit{utility}$^{\ast }$ profiles. i.
e.,} \textit{the North East frontier of }$\Psi (\mathcal{Q})$ \textit{if we
divide goods, and its SouthWest frontier if we divide bads.}\smallskip

The structure of efficient allocations in the linear domain is key to
several of our results. Given $z\in \Phi (N,A)$ we define the $N\times A$%
-bipartite \textit{consumption graph} $\Gamma (z)=\{(i,a)|z_{ia}>0\}$%
.\smallskip

\noindent \textbf{Lemma 1}

\noindent $a)$\textit{\ Fix a problem} $\mathcal{Q}=(N,A,u)$. \textit{If }$%
U\in \Psi ^{eff}(\mathcal{Q})$\textit{\ there is some }$z\in \Phi (N,A)$%
\textit{\ representing }$U$\textit{\ such that }$\Gamma (z)$\textit{\ is a
forest (an acyclic graph). For such allocation }$z$ \textit{the matrix }$%
[z_{ia}]$\textit{\ has at least }$(n-1)(p-1)$\textit{\ zeros.\smallskip }

\noindent $b)$\textbf{\ }\textit{Fixing }$N,A$\textit{, on an open dense
subset }$\mathcal{U}^{\ast }(N,A)$\textit{\ of matrices }$u\in 
\mathbb{R}
_{+}^{N\times A}$,\textit{\ every efficient \textit{utility}$^{\ast }$
profile }$U\in \Psi ^{eff}(N,A,u)$\textit{\ is achieved by a single
allocation }$z$\textit{.}

\textit{See in Subsection 1 the definition of }$\mathcal{U}^{\ast }(N,A)$%
\textit{\ and the proof of Lemma 1.\smallskip }

We use two equivalent definitions of a division rule, in terms of \textit{%
\textit{utility}}$^{\ast }$ profiles, or of feasible allocations. As this
will cause no confusion, we use the \textquotedblleft division
rule\textquotedblright\ terminology in both cases. When we rescale each $%
u_{i}$ as $\lambda _{i}u_{i}$ the new profile is written $\lambda \ast u$%
.\smallskip

\noindent \textbf{Definition 2}

\noindent $i)$ \textit{A division rule }$F$\textit{\ associates to every
problem }$\mathcal{Q}=(N,A,u)$\textit{\ a set of \textit{utility}}$^{\ast }$%
\textit{\ profiles }$F(\mathcal{Q})\subset \Psi (\mathcal{Q})$\textit{.
Moreover }$F(N,A,\lambda \ast u)=\lambda \ast F(N,A,\lambda \ast u)$\textit{%
\ for any rescaling }$\lambda $\textit{\ with }$\lambda _{i}>0$\textit{\ for
all }$i$\textit{.}

\noindent $ii)$\textit{\ A division rule }$f$\textit{\ associates to every
problem }$\mathcal{Q}=(N,A,u)$\textit{\ a subset }$f(\mathcal{Q})$\textit{\
of }$\Phi (N,A)$\textit{\ such that for any }$z,z^{\prime }\in \Phi (N,A)$%
\textit{:}%
\begin{equation}
\{z\in f(\mathcal{Q})\text{ and }u_{i}\cdot z_{i}=u_{i}\cdot z_{i}^{\prime }%
\text{ for all }i\in N\}\Longrightarrow z^{\prime }\in f(\mathcal{Q})
\label{17}
\end{equation}%
\textit{Moreover }$f(N,A,\lambda \ast u)=f(\mathcal{Q})$\textit{\ for any
rescaling }$\lambda $\textit{\ where }$\lambda _{i}>0$\textit{\ for all }$i$%
.\smallskip

The one-to-one mapping from $F$ to $f$ is clear. Definition 2 makes no
distinction between two allocations with identical welfare consequences, a
property often called \textit{Pareto-Indifference}.

We speak of a \textit{single-valued} division rule if $F(\mathcal{Q})$ is a
singleton for all $\mathcal{Q}$, otherwise the rule is \textit{multi-valued}%
. Single-valued rules are much more appealing, as they eschew the further
negotiation required to converge on a single division.

\section{Two division rules}

The definition of the \textit{Egalitarian }rule goes back to Pazner and
Schmeidler (\cite{PS}), who introduced it as a welfarist alternative to the
competitive approach. In our context we first normalize utilities$^{\ast }$
so that eating the entire pile of items (goods or bads) gives a utility$%
^{\ast }$ of $1$ to each participant, then find an efficient utility profile
where normalized utilities$^{\ast }$ are equal.

We call a problem $\mathcal{Q}$ in Definition 1 \textit{normalized} if $%
u_{i}\cdot e^{A}=1$ for all $i$. Because division rules are invariant to
rescaling, it is enough to define such a rule $F$ on the subdomain of
normalized division problems: if $\mathcal{Q}=(N,A,u)$ is not normalized we
simply set $F(\mathcal{Q})=F(N,A,\widetilde{u})$ where $\widetilde{u}_{i}=%
\frac{1}{u_{i}\cdot e^{A}}u_{i}$ for all $i$.

Interestingly the definition of the Egalitarian rule is simpler when we
divide bads rather than goods. In the case of bads there is always a
(unique) efficient normalized utility profile such that $%
U_{i}^{eg}=U_{j}^{eg}$ for all $i,j$, which the rule selects. Not so in the
case of goods. Consider for instance three agents and two goods $a,b$ where
agent $1$ likes only $a$ while agents $2,3$ like only $b$. Efficiency
implies that agent $1$ eats $a$ and $U_{1}=1$, while at least one of $2,3$
gets $U_{i}\leq \frac{1}{2}$. In this example the Egalitarian rule naturally
splits $b$ equally between $2$ and $3$.

In the case of goods we must use a familiar ordering of utility profiles.
For any $U\in 
\mathbb{R}
_{+}^{N}$, let $U^{\ast }\in 
\mathbb{R}
_{+}^{n}$ be the vector with the same coordinates arranged increasingly, and
recall that the \textit{leximin ordering} $\succeq _{lx}$compares $U$ and $%
U^{\prime }$ as the lexicographic ordering of $%
\mathbb{R}
_{+}^{n}$ compares $U^{\ast }$ and $U^{\prime \ast }$. The leximin ordering
has a unique maximum on every convex compact of $%
\mathbb{R}
_{+}^{N}$ (see e.g. \cite{M0}).\smallskip

\noindent \textbf{Definition 3 }\textit{Fix a normalized problem }$\mathcal{Q%
}=(N,A,u)$.

\noindent $i)$ \textit{If we divide goods,the Egalitarian division rule }$%
F^{eg}$\textit{\ picks the utility profile }$U^{eg}$ \textit{maximizing the l%
}ex\textit{imin ordering in} $\Psi (\mathcal{Q})$.

\noindent $ii)$ \textit{If we divide bads,the Egalitarian division rule }$%
F^{eg}$\textit{\ picks the efficient utility profile }$U^{eg}$ \textit{such
that }$U_{i}^{eg}=U_{j}^{eg}$ for all $i,j$.\smallskip

We check that the Definition $ii)$ makes sense. Set $\theta =\min_{\Psi (%
\mathcal{Q})}\max_{i}U_{i}$ and pick  $\overline{U}$ in $\Psi (\mathcal{Q})$
achieving $\theta $. Note that $\theta $ is positive. Suppose $\overline{U}%
_{1}<\theta $: then for any $i\geq 2$ such that $u_{i}\cdot z_{i}=\theta $
we take a small amount of some $a$ such that $u_{ia}>0$ and $z_{ia}>0$, and
give it to agent $1$. If these amounts are small enough, we get an
allocation $z^{\prime }$ where $u_{i}\cdot z_{i}^{\prime }<\theta $ for all $%
i$, including $1$, contradicting the definition of $\theta $. Thus $%
\overline{U}_{i}=\theta $ for all $i$. \ Now check that $\overline{U}$ is
efficient by a similar argument: if there is some $z\in \Phi (N,A)$ such
that $u_{i}\cdot z_{i}\leq \theta $ for all $i$ and $u_{1}\cdot z_{1}<\theta 
$, we can transfer some bads from any agent $i$ such that $u_{i}\cdot
z_{i}=\theta $ to agent $1$, and contradict again the definition of $\theta $%
.\smallskip 

\noindent \textbf{Definition 4 }\textit{Fix a problem }$\mathcal{Q}=(N,A,u)$.

\noindent $i)$\textit{\ If we divide goods} \textit{we call the allocation }$%
z\in \Phi (N,A)$\textit{\ competitive if there is a price }$p\in 
\mathbb{R}
_{+}^{A}$\textit{\ such that }$\sum_{A}p_{a}=n$\textit{\ and}%
\begin{equation}
z_{i}\in \arg \max_{y_{i}\in 
\mathbb{R}
_{+}^{A}}\{u_{i}\cdot y_{i}|p\cdot y_{i}\leq 1\}\text{ for all }i  \label{27}
\end{equation}%
$ii)$ \textit{If we divide bads} \textit{we call the allocation }$z\in \Phi
(N,A)$\textit{\ competitive if there is a price }$p\in 
\mathbb{R}
_{+}^{A}$\textit{\ such that }$\sum_{A}p_{a}=n$\textit{\ and}%
\begin{equation}
z_{i}\in \arg \min_{y_{i}\in 
\mathbb{R}
_{+}^{A}}\{u_{i}\cdot y_{i}|p\cdot y_{i}\geq 1\}\text{ for all }i  \label{41}
\end{equation}%
\textit{and for all }$a\in A$%
\begin{equation}
p_{a}=0\text{ if }u_{ia}=0\text{ for some }i\in N  \label{49}
\end{equation}

In the case of goods this Definition implies $U\gg 0$ because each row $%
u_{i} $ is non null and agent $i$ can afford some amount\textit{\ }of a good
he likes. Inequality $U\gg 0$ holds for bads as well because agent $i$ must
buy some bads with a positive price, and by~(\ref{49}) he dislikes such bads.

We write the Competitive rule as $f^{c},F^{c}$: it selects all competitive
allocations or utility$^{\ast }$ profiles. Existence of such allocations
both for goods and for bads is well known, as explained in the companion
paper \cite{BMSY}.

Property~(\ref{49}) rules out inefficient solutions of system~(\ref{41}).
For example assume two bads, two agents and%
\begin{equation*}
\begin{array}{ccc}
& a & b \\ 
u_{1} & 2 & 1 \\ 
u_{2} & 0 & 1%
\end{array}%
\end{equation*}%
There are three solutions of~(\ref{41})%
\begin{equation*}
\begin{array}{ccc}
& a & b \\ 
z_{1} & 1/4 & 1 \\ 
z_{2} & 3/4 & 0 \\ 
p & 4/3 & 2/3%
\end{array}%
\text{ \ \ ; \ }%
\begin{array}{ccc}
& a & b \\ 
z_{1} & 0 & 1/2 \\ 
z_{2} & 1 & 1/2 \\ 
p & 0 & 2%
\end{array}%
\text{ \ \ ; \ \ }%
\begin{array}{ccc}
& a & b \\ 
z_{1} & 0 & 1 \\ 
z_{2} & 1 & 0 \\ 
p & 1 & 1%
\end{array}%
\end{equation*}%
The left one is inefficient, and~(\ref{49}) rules out the right one (though
it is efficient).\smallskip

We give two additional characterizations of competitive allocations,
critical to most of our results. The first one is a simple and intuitive
system of inequalities.\smallskip

\noindent \textbf{Lemma 2} \textit{Fix a problem }$\mathcal{Q}=(N,A,u)$%
\textit{. Then }$U\in F^{c}(\mathcal{Q})$\textit{\ if and only if:}

\noindent $i)$\textit{\ Case of goods: }$U\gg 0$ and $U=(u_{i}\cdot
z_{i})_{i\in N}$\textit{\ for some }$z\in \Phi (N,A)$\textit{\ such that for
all }$i\in N$%
\begin{equation}
\text{for all }a\in A\text{: }z_{ia}>0\Longrightarrow {\large \{}\frac{u_{ia}%
}{U_{i}}\geq \frac{u_{ja}}{U_{j}}\text{ for all }j\in N{\large \}}
\label{42}
\end{equation}

\noindent $ii)$\textit{\ Case of bads: }$U\gg 0$ and $U=(u_{i}\cdot
z_{i})_{i\in N}$\textit{\ for some }$z\in \Phi (N,A)$\textit{\ such that for
all }$i\in N$%
\begin{equation}
\text{for all }a\in A\text{: }z_{ia}>0\Longrightarrow {\large \{}\frac{u_{ia}%
}{U_{i}}\leq \frac{u_{ja}}{U_{j}}\text{ for all }j\in N{\large \}}
\label{43}
\end{equation}

The next result is a geometric representation of competitive allocations.
Given $\mathcal{Q}$ we call $U$ a \textit{critical point of the Nash product}
$\mathcal{N}(U)={\LARGE \Pi }_{i\in N}U_{i}$ \textit{in} $\Psi (\mathcal{Q})$
if $U\in \Psi (\mathcal{Q}),U\gg 0$, and the hyperplane supporting the upper
contour of $\mathcal{N}$ at $U$ supports $\Psi (\mathcal{Q})$ as well. Such
critical points include the strictly positive local maxima and local minima
of $\mathcal{N}$ in $\Psi (\mathcal{Q})$.\smallskip

\noindent \textbf{Proposition 1}\textit{\ Fix a problem }$\mathcal{Q}%
=(N,A,u) $.

\noindent $i)$ \textit{If we divide goods the Competitive utility profile }$%
F^{c}(\mathcal{Q)}$ \textit{is the unique maximizer of the Nash product in }$%
\Psi (\mathcal{Q})$.

\noindent $ii)$ \textit{If we divide bads} \textit{the Competitive
disutility profiles in }$F^{c}(\mathcal{Q)}$\textit{\ are exactly all the
critical points of the Nash product }$\mathcal{N}$ \textit{in }$\Psi ^{eff}(%
\mathcal{Q})$.$\smallskip $

Statement $i)$ is well known and goes back to Eisenberg and Gale (\cite{EG}%
). A more general version of statement $ii)$ is proven in \cite{BMSY}, to
which we refer the reader.

A consequence of Definition 3 and Proposition 1 is that both rules $%
F^{eg},F^{c}$ are \textquotedblleft welfarist\textquotedblright , in the
sense that the utility profiles they choose are entirely determined by the
set of feasible utilities.

\section{Comparing the two rules}

\subsection{Fair Share Guarantee}

The mild and compelling test known as \textit{Fair Share Guarantee} already
appears in the early cake division literature. The idea is to take the
possibly inefficient equal division of the resources (where each agent
receives $\frac{1}{n}e^{A}$) as the default option that each participant can
(virtually) enforce, thus setting a lower bound on individual welfare.

\textbf{Fair Share Guarantee }(FSG) for any $\mathcal{Q}$ and any $U\in F(%
\mathcal{Q})$ we have%
\begin{equation*}
U_{i}\geq \frac{1}{n}u_{i}\cdot e^{A}\text{ (goods); }U_{i}\leq \frac{1}{n}%
u_{i}\cdot e^{A}\text{ (bads)}
\end{equation*}%
Clearly both rules $F^{eg}$ and $F^{c}$meet FSG. However only the former
meets the following strict version of FSG.

\textbf{Strict} \textbf{Fair Share Guarantee }(SFSG) if in $\mathcal{Q}$ the
equal split allocation is efficient, then $U_{i}=\frac{1}{n}u_{i}\cdot e^{A}$
for all $i$; if it is not we have, for any $U\in F(\mathcal{Q})$%
\begin{equation*}
U_{i}>\frac{1}{n}u_{i}\cdot e^{A}\text{ (goods); }U_{i}<\frac{1}{n}%
u_{i}\cdot e^{A}\text{ (bads)}
\end{equation*}

\noindent \textbf{Lemma 3 }\textit{The Egalitarian rule} \textit{meets SFSG;
the Competitive rule does not if we have at least two agents and/or at least
two items.\smallskip }

The SFSG test reveals a weakness of the Competitive rule that the
Egalitarian rule does not share.

That the Egalitarian rule meets SFSG is clear if we divide bads, by
statement $ii)$ in Definition 3. In order to prove it when we divide goods,
fix a problem $\mathcal{Q}$ and partition the agents as follows%
\begin{equation*}
U_{i}^{eg}=\frac{1}{n}u_{i}\cdot e^{A}\text{ for }i\in N_{0}\text{ ; }%
U_{i}^{eg}>\frac{1}{n}u_{i}\cdot e^{A}\text{ for }i\in N_{+}
\end{equation*}%
If equal split is not efficient the set $N_{+}$ is non empty; suppose $N_{0}$
is non empty as well and derive a contradiction. Take any good $a$ and agent 
$i$ in $N_{+}$ such that $z_{ia}>0$ and $u_{ia}>0$: if some $j$ in $N_{0}$
likes $a$ as well, we can improve the utility profile for the leximin
ordering by transferring a little $a$ from $i$ to $j$. Therefore nobody in $%
N_{0}$ likes any of the goods eaten by $N_{+}$, so by splitting equally the
goods that agents in $N_{0}$ like among themselves, they each get $\frac{1}{%
|N_{0}|}u_{i}\cdot e^{A}$ which is the desired contradiction.

\begin{figure}[h!]
\centering
{\includegraphics[width=10cm, clip=true, trim=0cm 6cm 0cm 5.5cm]{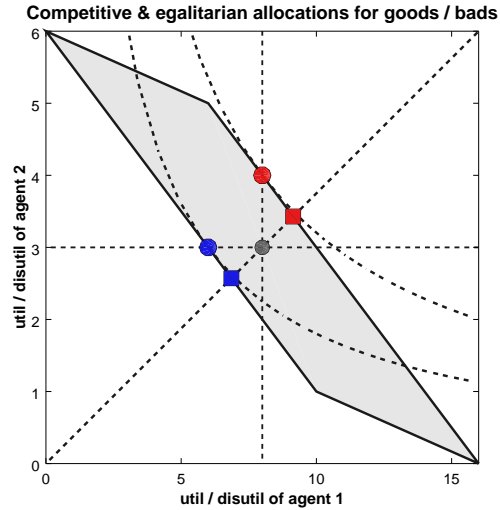}}
\caption{For the first example with goods the red circle and the red square represent the competitive and egalitarian utility profiles, respectively. Gray point is the equal split utility profile. Blue points are for the case of bads.}
\label{fig1}
\end{figure}

We illustrate the second statement by two examples. In the first one we have
two agents and two items; the Competitive division is unique, whether the
matrix $u$ represent utilities for goods or disutilities for bads:%
\begin{equation*}
\begin{array}{ccc}
& a & b \\ 
u_{1} & 10 & 6 \\ 
u_{2} & 5 & 1%
\end{array}%
\text{ \ for goods: }z^{c}=%
\begin{array}{ccc}
& a & b \\ 
z_{1} & 1/5 & 1 \\ 
z_{2} & 4/5 & 0 \\ 
p & 5/4 & 3/4%
\end{array}%
\text{ \ for bads: }z^{c}=%
\begin{array}{ccc}
& a & b \\ 
z_{1} & 3/5 & 0 \\ 
z_{2} & 2/5 & 1 \\ 
p & 5/3 & 1/3%
\end{array}%
\end{equation*}%
The corresponding utilities and disutilities, illustrated in Figure~\ref{fig1}, are:
for goods: $U^{c}=(8,4)$, $U^{eg}=(9\frac{1}{7},3\frac{3}{7})$; and for
bads: $U^{c}=(6,3)$, $U^{eg}=(6\frac{6}{7},2\frac{4}{7})$.

Our second example is a problem with goods with $n$ agents and $(n-1)$
goods, where $F^{c}$ violates SFSG. In this canonical problem the contrast
between our two rules is especially stark. We have $n$ agents and $(n-1)$
goods. The first $(n-1)$ agents are \textit{single-minded}: agent $i,1\leq
i\leq n-1,$ likes only good $a_{i}$. The last agent $n$ is \textit{flexible}%
, he likes all goods equally%
\begin{equation*}
\begin{array}{ccccc}
\text{good} & a_{1} & a_{2} & \cdots & a_{n-1} \\ 
u_{1} & 1 & 0 & 0 & 0 \\ 
u_{2} & 0 & 1 & 0 & 0 \\ 
\cdots & 0 & 0 & 1 & 0 \\ 
u_{n-1} & 0 & 0 & 0 & 1 \\ 
u_{n} & 1 & 1 & 1 & 1%
\end{array}%
\end{equation*}%
The competitive price is $\frac{n}{n-1}$ for every good: each single-minded
agent buys $\frac{n-1}{n}$ units of \textquotedblleft his\textquotedblright\
good while the flexible agent $n$ eats $\frac{1}{n}$-th of each good%
\begin{equation}
z_{i}=\frac{n-1}{n}e^{a_{i}}\text{ for }1\leq i\leq n-1\text{ ; }z_{n}=\frac{%
1}{n}e^{A}  \label{47}
\end{equation}%
This is tough on the flexible agent who gets his fair share and no more,
while everybody else gets $(n-1)$ times more! The reason is that in this
example the Competitive allocation is the only one in the core from equal
split: the coalition of all single-minded agents does not need agent $n$ to
achieve its competitive surplus.

The Egalitarian rule splits each good $i$ equally between agent $i$ and
agent $n$, and everyone ends up with a share worth one half of the entire
manna, much above her Fair Share.%
\begin{equation*}
z_{i}=\frac{1}{2}e^{a_{i}}\text{ for }1\leq i\leq n-1\text{ ; }z_{n}=\frac{1%
}{2}e^{A}
\end{equation*}%
Here we submit that the Egalitarian allocation above gives \textit{too much}
to agent $n$, who gets (much) more than his fair share of \textit{every}
good. By contrast at a Competitive allocation, \textit{here and always},
everyone gets \textit{at most} a $\frac{1}{n}$-th share of \textit{at least}
one good: for all $i$ we have $\min_{a\in A}z_{ia}\leq \frac{1}{n}$.%
\footnote{%
If $z_{ia}>\frac{1}{n}$ for all $a$ the competitive price must be parallel
to $u_{i}$ (or eating of each good would not be a competitive demand) and
the equal budget condition $p\cdot z_{i}=p\cdot (\frac{1}{n}e^{A})$ gives $%
u_{i}\cdot z_{i}=u_{i}\cdot (\frac{1}{n}e^{A})$, contradiction. If we divide
bads every Competitive allocation satisfies similarly $\max_{a\in
A}z_{ia}\geq \frac{1}{n}$.}

For fixed sizes of $N$ and $A$ it would be interesting to understand at
which problems the $\ell _{\infty }$ or $\ell _{1}$ distance between the
profiles of normalized utilities at the competitive and egalitarian
allocations, is the largest possible. The canonical example may be a step
toward the answer.

\subsection{Resource Monotonicity}

More goods or fewer bads to divide should be good news (at least weakly) for
everyone: all agents \textquotedblleft own\textquotedblright\ the goods/bads
equally and welfare should be comonotonic to ownership. This general
solidarity property has played a major role in the modern fair division
literature: \cite{MT}, \cite{T1}. When it fails someone has an incentive to
sabotage the discovery of additional goods.

In the following definition $B$ is a subset of $A$ and we write $u_{[B]}$
for the restriction to $%
\mathbb{R}
_{+}^{N\times B}$ of the utility$^{\ast }$ matrix $u\in 
\mathbb{R}
_{+}^{N\times A}$.

\textbf{Resource Monotonicity}\textit{\ }(RM): for any two problems $%
\mathcal{Q=}(N,A,u)$ and $\mathcal{Q}^{\prime }\mathcal{=}(N,B,u_{[B]})$
where $B\subset A$, we have

\begin{equation}
\text{for any }U\in F(\mathcal{Q}),U^{\prime }\in F(\mathcal{Q}^{\prime })%
\text{: }U^{\prime }\leq U\text{ (goods) ; }U\leq U^{\prime }\text{ (bads)}
\label{19}
\end{equation}%
(going from $A$ to $B$ is bad news if we deal with goods, and good news with
bads).\smallskip

\noindent \textbf{Proposition 2}

\noindent $i)$ \textit{If we divide goods, the Competitive\ rule is Resource
Monotonic. For three or more agents the Egalitarian rule is not.}

\noindent $ii)$\textit{\ If we divide bads among three or more agents and
two or more bads, no efficient rule can meet Resource Monotonicity and Fair
Share Guarantee.\footnote{%
Note that we sketch the proof of statement $ii)$ in Subsection 7.2 of \cite%
{BMSY}. We provide here the complete proof in Subsection 6.3.}\smallskip }

Recall from \cite{MT} (see also \cite{TK}) that in any domain containing the
Leontief preferences, we cannot divide \textit{goods} efficiently while
ensuring FSG and RM (this is true even with two agents and two goods). This
makes the contrast of goods versus bads in the additive domain all the more
intriguing.

Proposition 2 gives a strong argument in support of the Competitive rule
when we divide goods, but also shows that this advantage disappears when we
divide bads.

The proof that $F^{c}$ meets RM for goods is in Subsection 6.3. We give here
a three agent example showing that $F^{eg}$ fails RM: this example
illustrates well the logic of the Egalitarian division of goods. It is easy
to show that $F^{eg}$ meets RM in two agent problems.

We compare the two problems below with $B=\{a,b,c\}$ and $A=\{a,b,c,d\}$%
\begin{equation*}
\mathcal{Q}(B)=%
\begin{tabular}{cccc}
& $a$ & $b$ & $c$ \\ 
$u_{1}$ & $3$ & $1$ & $1$ \\ 
$u_{2}$ & $1$ & $3$ & $1$ \\ 
$u_{3}$ & $1$ & $1$ & $3$%
\end{tabular}%
\text{ ; }\mathcal{Q}(A)=%
\begin{tabular}{ccccc}
& $a$ & $b$ & $c$ & $d$ \\ 
$u_{1}$ & $3$ & $1$ & $1$ & $0$ \\ 
$u_{2}$ & $1$ & $3$ & $1$ & $4$ \\ 
$u_{3}$ & $1$ & $1$ & $3$ & $4$%
\end{tabular}%
\end{equation*}

Problem $\mathcal{Q}(B)$ is symmetric. Any efficient and symmetric rule
allocates goods \textquotedblleft diagonally\textquotedblright : agent $1$
gets all of $a$ and so on; normalized utilities are $\frac{3}{5}$. In $%
\mathcal{Q}(A)$ the natural idea is to keep the same allocation of $a,b,c$
and divide $d$ equally between agents $2$ and $3$, because agent $1$ does
not care for $d$. This is what $F^{c}$ recommends (prices are $(1,\frac{3}{5}%
,\frac{3}{5},\frac{4}{5})$). But the normalized utilities at this allocation
are $(\frac{3}{5},\frac{5}{9},\frac{5}{9})$, so the $F^{eg}$ must compensate
agents $2,3$ for the \textit{loss }in normalized utilities caused by the 
\textit{gain} of some new good! Equality is restored at the allocation%
\begin{equation*}
z^{eg}=%
\begin{tabular}{cccc}
$a$ & $b$ & $c$ & $d$ \\ 
$55/59$ & $0$ & $0$ & $0$ \\ 
$2/59$ & $1$ & $0$ & $1/2$ \\ 
$2/59$ & $0$ & $1$ & $1/2$%
\end{tabular}%
\end{equation*}

\noindent where agent $1$'s welfare has decreased.

\subsection{Independence of lost bids and local misreports}

By a local misreport of one's preferences, we mean that a change of report
that does not affect the consumption forest $\Gamma (z)$ identified in Lemma
1. For example the utility matrix $u$ is in $\mathcal{U}^{\ast }(N,A)$
(Subsection 6.1) so that this forest does not change in a neighborhood of $u$%
.

We call agent $i$'s marginal utility $u_{ia}$ her \textquotedblleft
bid\textquotedblright\ for item $a$; given a problem $\mathcal{Q}=(N,A,u)$
we say that $i$'s bid is \textquotedblleft losing\textquotedblright\ if $%
z_{ia}=0$, and \textquotedblleft winning\textquotedblright\ if if $z_{ia}=1$.

Under the Egalitarian rule dividing goods, a profitable local manipulation
is to \textit{inflate} my losing bids and \textit{deflate} my winning bids;
if we divide bads, I want instead to inflate my winning bids and deflate the
losing ones. This is the familiar bargaining tactic of playing down the
worth on your concession (my share) while exaggerating that of my concession
(your share). For instance in problem $\mathcal{Q}(A)$ of the previous
section, if agent $1$ reports $u_{1a}^{\prime }=\frac{5}{2}$ instead of $%
u_{1a}=3$ (or increases both $u_{1b}^{\prime }$ ans $u_{1c}^{\prime }$ to $%
\frac{6}{5}$) the apparent egalitarian allocation $z^{\prime }$ becomes the
competitive allocation (at $u$) where she eats all $a$.

Given a problem $\mathcal{Q=(}N,A,u\mathcal{)}$ with goods with a single
egalitarian allocation $z=f^{eg}(u)$, we call $u_{i}^{\prime }$ a \textit{%
simple misreport }by agent $i$ if we have for all $a\in A$%
\begin{equation}
z_{ia}=0\Longrightarrow u_{ia}^{\prime }\geq u_{ia}\text{ ; }%
z_{ia}=1\Longrightarrow u_{ia}^{\prime }\leq u_{ia}\text{ ; }%
0<z_{ia}<1\Longrightarrow u_{ia}^{\prime }=u_{ia}  \label{22}
\end{equation}%
and at least one inequality is strict. In a problem $\mathcal{Q}$ with bads,
the same definition applies upon changing the sign of the
inequalities.\smallskip

\noindent \textbf{Lemma 4 }

\textit{Fix} $\mathcal{Q}$ \textit{and a simple misreport }$u_{i}^{\prime }$%
\textit{\ as above. If }$f^{eg}(u_{i}^{\prime },u_{-i})=z^{\prime }$\textit{%
, and }$\Gamma (z^{\prime })=\Gamma (z)$ \textit{the misreport is profitable}%
\begin{equation*}
\text{goods: }u_{i}\cdot z_{i}^{\prime }>u_{i}\cdot z_{i}\text{ ; bads: }%
u_{i}\cdot z_{i}^{\prime }<u_{i}\cdot z_{i}
\end{equation*}

The impact of local misreports on competitive allocations is very different.
Lowering or raising a losing bid has no impact at all on the entire
allocation. This is clear from Lemma 2. Fix a goods problem $\mathcal{Q}$,
with $z=f^{c}(u)$ and the competitive price $p$. Then the utility
maximization~(\ref{27}) implies $\frac{u_{ia}}{U_{i}}\leq p_{a}$ for all $a$%
, with equality if $a\in \lbrack z_{i}]$. Therefore if $i$ changes her
losing bid $u_{ia}$ to any $u_{ia}^{\prime }$ such that $\frac{%
u_{ia}^{\prime }}{U_{i}}\leq p_{a}$, the allocation $z$ is still in $%
f^{c}(u_{i}^{\prime },u_{-i})$ (system~(\ref{42}) still holds) so the
misreport is of no consequence to anybody. The same argument applies if we
divide bads: if $z\in f^{c}(u)$ and $i$ changes her losing bid $u_{ia}$ to $%
u_{ia}^{\prime }$, the allocation $z$ remains in $f^{c}(u_{i}^{\prime
},u_{-i})$ as long as inequality $\frac{u_{ia}^{\prime }}{U_{i}}\geq p_{a}$
is preserved.

There is an incentive aspect to this invariance property:\ to misreport on
an item that I do not end up consuming is \textquotedblleft
cheap\textquotedblright , because it is presumably harder to verify \textit{%
ex post} my marginal utility$^{\ast }$ for that item than for an item I am
actually eating.

Of course the Competitive rule is manipulable by misreporting winning bids,
whether we divide goods or bads. But unlike for the Egalitarian rule,
whether a profitable manipulation is to increase or decrease a winning bid
does depend upon the entire utility matrix.For instance in the three agents,
two goods problem%
\begin{equation*}
\begin{array}{ccc}
& a & b \\ 
u_{1} & 3 & 1 \\ 
u_{2} & \alpha & 1 \\ 
u_{3} & 1 & 3%
\end{array}%
\text{ where }\frac{1}{2}<\alpha <2\text{ }\Longrightarrow z^{c}=%
\begin{array}{ccc}
& a & b \\ 
z_{1} & (1+\alpha ^{-1})/3 & 0 \\ 
z_{2} & (2-\alpha ^{-1})/3 & (2-\alpha ^{-1})/3 \\ 
z_{3} & 0 & (1+\alpha ^{-1})/3%
\end{array}%
\end{equation*}%
the optimal misreport of $u_{2a}=\alpha $ is $u_{2a}^{\prime }=\sqrt{\alpha }
$, hence it can be above or below the true winning bid $\alpha $.\smallskip

\noindent \textbf{Definition 5 }\textit{The rule }$f$ \textit{is Independent
of Lost Bids (ILB) if for any two problems }$\mathcal{Q},\mathcal{Q}^{\prime
}$ on $N,A$\textit{\ where }$u,u^{\prime }$ \textit{differ only in the entry 
}$ia$,\textit{\ and such that }$u_{ia}>u_{ia}^{\prime }$ \textit{(goods) or} 
$u_{ia}<u_{ia}^{\prime }$ \textit{(bads), we have}%
\begin{equation}
\forall z\in f(\mathcal{Q}):z_{ia}=0\Longrightarrow z\in f(\mathcal{Q}%
^{\prime })  \label{24}
\end{equation}

\textbf{Equal Treatment of Equals} (ETE) is the familiar requirement that
the rule $F$ should not discriminate between two agents with identical
characteristics. For all $\mathcal{Q}$ and $i,j\in N$%
\begin{equation*}
u_{i}=u_{j}\Longrightarrow U_{i}=U_{j}\text{ for all }U\in F(\mathcal{Q})
\end{equation*}

\noindent \textbf{Proposition 3 }\textit{(goods or bads)}

\noindent \textit{If a division rule meets Efficiency, Independence of Lost
Bids, and at least one of Equal Treatment of Equals and Fair Share
Guaranteed, it contains the Competitive rule.\smallskip }

The Competitive rule for goods is characterized by adding single-valuedness
to the above requirements.

We show after the proof (Subsection 6.5) that ILB is a strictly weaker
requirement than Maskin Monotonicity (\cite{Ma}) in the linear domain. This
relates Proposition 3 to earlier results mentioned in Section 2.\smallskip

We discuss\textit{\ }the tightness of our characterization.

\noindent \textit{Drop ETE and FSG}. An asymmetric variant of the
competitive solution, where fixed income shares replace equal income shares,
is efficient and meets ILB. See Remark 1 in Section 3 of \cite{BMSY}.

\noindent \textit{Drop ILB. }The Egalitarian rule meets the other three
axioms.

\noindent \textit{Drop Efficiency}. It is tempting to consider the simple
Equal Split rule, dividing all items equally irrespective of preferences.
However by Definition 2 we must also select all allocations generating the
same utility profile, and this correspondence violates ILB. But a
constrained version of the Competitive rule, where we restrict individual
consumptions as in the assignment model of \cite{HZ} (i. e., $\sum_{A}z_{ia}=%
\frac{1}{n}|A|$ for all $i$), satisfies ETE and ILB by the same reasoning as
in unconstrained problems.

\subsection{Multiple Competitive divisions}

When we divide goods, both rules pick a unique utility profile, and both are
easy to compute, respectively by a linear or convex optimization program.
This remains true for the Egalitarian rule dividing bads, but not for the
Competitive rule.

\begin{figure}[h!]
\centering
{\includegraphics[width=10cm, clip=true, trim=0cm 6cm 0cm 5.5cm]{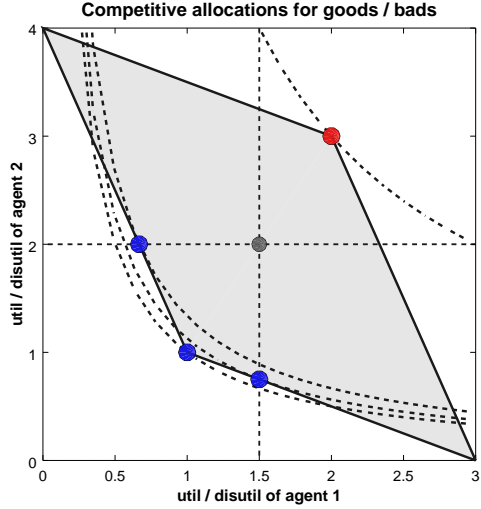}}
\caption{The unique competitive utility profile for goods (red) and multiple profiles for bads~(blue). The gray dot corresponds to the equal split utility profile.}
\label{fig2}
\end{figure}

In all numerical examples of bads problems discussed so far (in Section 4
and subsections 5.1, 5.2) the Competitive rule was single valued. The
simplest illustration of the unpalatable multiplicity issue has two agents
and two bads:%
\begin{equation*}
\begin{array}{ccc}
& a & b \\ 
u_{1} & 1 & 2 \\ 
u_{2} & 3 & 1%
\end{array}%
\text{ }\quad z^{c1}=%
\begin{array}{ccc}
& a & b \\ 
z_{1} & 1 & 1/4 \\ 
z_{2} & 0 & 3/4 \\ 
p & 2/3 & 4/3%
\end{array}%
\text{ }\quad z^{c2}=%
\begin{array}{ccc}
& a & b \\ 
z_{1} & 1 & 0 \\ 
z_{2} & 0 & 1 \\ 
p & 1 & 1%
\end{array}%
\text{ } \quad z^{c3}=%
\begin{array}{ccc}
& a & b \\ 
z_{1} & 2/3 & 0 \\ 
z_{2} & 1/3 & 1 \\ 
p & 3/2 & 1/2%
\end{array}%
\end{equation*}%
See Figure~\ref{fig2}. Note that at $z^{c1}$ agent $1$ gets only his Fair Share
utility level, while agent $2$ grabs all the surplus above equal split; at $%
z^{c3}$ agents $1$ and $2$ exchange roles.

Our second example has the same structure as the canonical example
concluding Subsection 5.1. We have $n$ agents, $(n-1)$ bads, and the first $%
(n-1)$ agents are single-minded, each over a different bad%
\begin{equation*}
\begin{array}{ccccc}
\text{bad} & a_{1} & a_{2} & \cdots & a_{n-1} \\ 
u_{1} & 1 & 3 & 3 & 3 \\ 
u_{2} & 3 & 1 & 3 & 3 \\ 
\cdots & 3 & 3 & 1 & 3 \\ 
u_{n-1} & 3 & 3 & 3 & 1 \\ 
u_{n} & 1 & 1 & 1 & 1%
\end{array}%
\end{equation*}

The allocation~(\ref{47}), respecting the symmetry between the first $n-1$
agents, is still competitive at the uniform price $\frac{n}{n-1}$ for each
bad: the flexible agent $n$ gets no relief from his equal split share.
However there are many more competitive divisions,\ all with different
utility profiles, and breaking at least partially the above symmetry.

Recall the notation $e^{S}$ for the vector in $%
\mathbb{R}
^{A}$ with $e_{i}^{S}=1$ if $i\in S$ and zero otherwise. Pick an integer $%
q,1\leq q\leq n-1$ and check that the allocation%
\begin{equation}
z_{i}=\frac{q}{q+1}e^{a_{i}}\text{ for }1\leq i\leq q\text{ ; }%
z_{j}=e^{a_{j}}\text{ for }q+1\leq j\leq n-1\text{ ; }z_{n}=\frac{1}{q+1}%
e^{\{a_{1},\cdots ,a_{q}\}}  \label{46}
\end{equation}%
is competitive for the prices $p_{a_{i}}=\frac{q+1}{q}$ for $1\leq i\leq q$
and $p_{a_{j}}=1$ for $q+1\leq j\leq n-1$. In particular agent $n$'s
disutility varies from $\frac{1}{2}$ to $\frac{n-1}{n}$. We could have
chosen any other subset of bads with size $q$, and any $q$ from $1$ to $n-1$%
. Thus there are $2^{n-1}-1$ different competitive allocations.

Note that by ILB (Definition 5) the entries $3$ in the disutility matrix
play almost no role in the computation above: replacing those entries by any 
$\gamma $ larger than $2$ does not affect the competitive divisions. By
contrast the (unique and symmetric) Egalitarian allocation depends heavily
upon $\gamma $: agent $n$'s disutility decreases to zero as $\gamma $ grows
to infinity.

Our first main result evaluates the extent of the multiplicity
issue.\smallskip

\noindent \textbf{Theorem 1 }\textit{For any problem }$\mathcal{Q}$ \textit{%
with bads}:

\noindent $i)$ \textit{The set }$F^{c}(\mathcal{Q})$\textit{\ of competitive
utility profiles is finite.}

\noindent $ii)$ \textit{For general }$n=|N|,m=|A|$,\textit{\ }$|F^{c}(%
\mathcal{Q})|$\textit{\ can be as high as }$2^{\min \{n,m\}}-1$\textit{\ if }%
$n\neq m$\textit{, and }$2^{n-1}-1$\textit{\ if} $n=m$.

\noindent $iii)$ For $n=2$ t\textit{he upper bound on }$|F^{c}(\mathcal{Q})|$%
\textit{\ is }$2m-1$.

\noindent $iv)$ For $m=2$ t\textit{he upper bound on }$|F^{c}(\mathcal{Q})|$%
\textit{\ is }$2n-1$.\textit{\smallskip }

There is a simple exponential upper bound on the number of distinct
Competitive allocations for general $n,m$. For a given consumption graph $%
\Gamma $, there is at most one competitive allocation (utilitywise): the
graph determines the set of active KKT constraints, and we can recover the
allocation from Lemma 2. Thus $|F^{c}(\mathcal{Q})|$ cannot exceed the
number of bipartite forests on $n+m$ vertices, which is bounded by $%
2^{(n+m)\ln (nm)}$.\footnote{%
Because we know that there are at least $m$ and at most $n+m-1$ edges in the
graph and there are $nm$ options to trace each edge.}

The canonical example before Theorem 1 proves half of statement $ii)$. The
longer proofs of statements $iii)$ and $iv)$ rely on the fact that for $n=2$
a problem is entirely described by the sequence of ratios $\frac{u_{1a}}{%
u_{2a}}$, and for $m=2$ by the sequence of ratios $\frac{u_{ia}}{u_{ib}}$.
This allows a closed form description of all competitive allocations.

A by-product of these proofs is that $|F^{c}(\mathcal{Q})|$ is odd on an
open dense subset of the problems where $n=2$ and/or $m=2$\textit{.\ }A very
plausible conjecture is that this is true as well for any $n,m$.

\subsection{Single-valued envy free rules: discontinuity}

In order to bypass the unpalatable multiplicity issue just discussed we must
identify a normatively appealing single-valued selection from the set of
Competitive divisions of bads. For instance if the problem involves only two
agents and/or two bads, the set of efficient and envy-free allocations is a
one-dimensional line (as explained in the proof of Theorem 1) with,
generically, an odd number of competitive allocations, so we can choose the
median allocation. Alternatively, in problems of any size we can pick among
efficient allocations the one maximizing the product of disutilities: it is
competitive and generically unique (Lemmas 3, 4 in Section 6 of \cite{BMSY}%
). But no such selection can meet the following compelling regularity
requirement.\smallskip

\textbf{Continuity }(CONT)\textit{\ }of the \textit{single-valued} division
rule $F$. For all $N$, $A$, the function $u\rightarrow F(N,A,u)$\ is
continuous in\textit{\ }$%
\mathbb{R}
_{++}^{N\times A}$\textit{.}\smallskip 

The Egalitarian rule to divide goods or bads, and the (single-valued)
Competitive rule to divide goods, are clearly continuous. But no selection
of the Competitive rule to divide bads is continuous. We deduce this
negative result from a much more general statement involving a familiar test
of fairness.\smallskip

\textbf{No Envy} (NE) of the possibly multivalued rule $f$. For all $%
\mathcal{Q}$ and all $i,j\in N$%
\begin{equation*}
\text{for any }z\in f(\mathcal{Q})\text{: }u^{i}\cdot z^{i}\geq u^{i}\cdot
z^{j}\text{ (goods); }u^{i}\cdot z^{i}\leq u^{i}\cdot z^{j}\text{ (bads)}
\end{equation*}

No Envy ensures equality of opportunity \textquotedblleft ex
post\textquotedblright\ (after we cut the cake), just like a competitive
allocation offers equality of opportunity \textquotedblleft ex
ante\textquotedblright\ (the common budget set). As is well known the set of
efficient and envy-free allocations contains much more than the competitive
ones. For instance with two agents, it contains all efficient allocations
guaranteeing Fair Shares. It is therefore surprising, and disappointing that
in the division of bads even this fairly permissive test is incompatible
with Continuity.\smallskip

\textbf{Theorem 2 }\textit{Say}\textbf{\ }\textit{we divide at least two
bads between at least four agents and fix a division rule }$f,F$. \textit{If 
}$F$\textit{\ is single-valued and Continuous, then }$f$ \textit{cannot be
also Efficient and Envy-Free.\smallskip }

The statement is tight. A single-valued selection of $F^{c}$ is such that
the corresponding selection of $f^{c}$ is Efficient and Envy-free. The rule $%
F^{eg}$ is Efficient and Continuous. Finally the rule derived from the equal
division of all bads, irrespective of disutility functions,\footnote{%
Defined by $f(\mathcal{Q})=\{z\in \Phi (N,A)|u_{i}\cdot z_{i}=\frac{1}{n}%
u_{i}\cdot e^{A}\}$, to meet Pareto indifference.} is Envy-Free and
Continuous.

We prove Theorem 2 as a Corollary of Proposition 4, our last result. There
we focus on the topological structure of the set $\mathcal{A}$ of efficient
and envy-free allocations in problems with two bads $a,b,$ and any number of
agents.\smallskip

\textbf{Proposition 4 }\textit{If we divide at least two bads between at
least three agents, there are problems }$\mathcal{Q}$ \textit{where the set }%
$\mathcal{A}$ \textit{of efficient and envy-free allocations, and the
corresponding set of disutility profiles, have }$\lfloor \frac{2n+1}{3}%
\rfloor $ \textit{connected components.\smallskip }

The set $\mathcal{A}$\ is clearly connected if all problems with two agents,
whether we divide goods or bads. In a general problem with \textit{goods},
we do not know if the set $\mathcal{A}$ is always connected.

\section{Appendix: Proofs}

\subsection{Lemma 1}

\subsubsection{$a)$ the consumption forest at efficient allocations}

\noindent Fix a problem $\mathcal{Q}$. To fix ideas we think of the items as
goods but the proof is identical for bads\textit{. }Pick $z$ representing $%
U\in \Psi ^{eff}(\mathcal{Q})$ and assume there is a $K$-cycle in $\Gamma
(z) $: $z_{ka_{k}},z_{ka_{k-1}}>0$ for $k=1,\cdots ,K$, where $a_{0}=a_{K}$.
Then $u_{ka_{k}},u_{ka_{k-1}}$ are positive for all $k$: if $u_{ka_{k}}=0$
efficiency and $\sum_{i\in N}u_{ia_{k}}>0$ imply $z_{ka_{k}}=0$.

\noindent Assume now%
\begin{equation}
\frac{u_{1a_{1}}}{u_{2a_{1}}}\cdot \frac{u_{2a_{2}}}{u_{3a_{2}}}\cdot \cdots
\cdot \frac{u_{(K-1)a_{K-1}}}{u_{Ka_{K-1}}}\cdot \frac{u_{Ka_{K}}}{u_{1a_{K}}%
}>1  \label{1}
\end{equation}%
Then we can pick arbitrarily small positive numbers $\varepsilon _{k}$ such
that%
\begin{equation}
\frac{u_{1a_{1}}\cdot \varepsilon _{1}}{u_{1a_{K}}\cdot \varepsilon _{K}}>1,%
\frac{u_{2a_{2}}\cdot \varepsilon _{2}}{u_{2a_{1}}\cdot \varepsilon _{1}}%
>1,\cdots ,\frac{u_{Ka_{K}}\cdot \varepsilon _{K}}{u_{Ka_{K-1}}\cdot
\varepsilon _{K-1}}>1  \label{25}
\end{equation}%
and the corresponding transfer to each agent $k$ of $\varepsilon _{k}$ units
of good $k$ against $\varepsilon _{k-1}$ units of good $k-1$ is a Pareto
improvement, contradiction. Therefore~(\ref{1}) is impossible; the opposite
strict inequality is similarly ruled out so we conclude%
\begin{equation}
\frac{u_{1a_{1}}}{u_{2a_{1}}}\cdot \frac{u_{2a_{2}}}{u_{3a_{2}}}\cdot \cdots
\cdot \frac{u_{(K-1)a_{K-1}}}{u_{Ka_{K-1}}}\cdot \frac{u_{Ka_{K}}}{u_{1a_{K}}%
}=1  \label{16}
\end{equation}%
Now if we perform a transfer as above where%
\begin{equation*}
\frac{u_{1a_{1}}\cdot \varepsilon _{1}}{u_{1a_{K}}\cdot \varepsilon _{K}}=%
\frac{u_{2a_{2}}\cdot \varepsilon _{2}}{u_{2a_{1}}\cdot \varepsilon _{1}}%
=\cdots =\frac{u_{Ka_{K}}\cdot \varepsilon _{K}}{u_{Ka_{K-1}}\cdot
\varepsilon _{K-1}}=1
\end{equation*}%
the utility profile $U$ is unchanged. If we choose the numbers $\varepsilon
_{k}$ as large as possible for feasibility, this will bring at least one
entry $(k,a_{k})$ or $(k,a_{k-1})$ to zero, so in our new representation $%
z^{\prime }$ of $U$ the graph $\Gamma (z^{\prime })$ has fewer edges. We can
clearly repeat this operation until we eliminate all cycles of $\Gamma (z)$.

The last statement follows at once from the fact that a forest with $n+m$
vertices contains at most $n+m-1$ edges.\smallskip

\subsubsection{$b)$ at almost all profiles each efficient utility profile is
achieved by a single allocation}

\noindent We let $\mathcal{U}^{\ast }(N,A)$ be the open and dense subset of $%
\mathbb{R}
_{+}^{N\times A}$ such that for any cycle $\mathcal{C=\{}1,a_{1},2,a_{2},%
\cdots ,a_{K},1\}$ in the complete bipartite graph $N\times A$ we have $\pi (%
\mathcal{C)}=\frac{u_{1a_{1}}}{u_{2a_{1}}}\cdot \frac{u_{2a_{2}}}{u_{3a_{2}}}%
\cdot \cdots \cdot \frac{u_{(K-1)a_{K-1}}}{u_{Ka_{K-1}}}\cdot \frac{%
u_{Ka_{K}}}{u_{1a_{K}}}\neq 1$ ((\ref{16}) fails) and moreover $u_{ia}>0$
for all $i,a$. It is clearly an open dense subset of $%
\mathbb{R}
_{+}^{N\times A}$.

We pick a problem $\mathcal{Q}\ $with $u\in 
\mathbb{R}
_{+}^{N\times A}$, fix $U\in \Psi ^{eff}(\mathcal{Q)}$ and assume there are
two different $z,z^{\prime }\in \Phi (N,A)$ such that $u\cdot z=u\cdot
z^{\prime }=U$. Pick a pair $1,a_{1}$ such that $z_{1a_{1}}>z_{1a_{1}}^{%
\prime }$. Because $a_{1}$ is eaten in full there is some agent $2$ such
that $z_{2a_{1}}<z_{2a_{1}}^{\prime }$ and because $u_{2}\cdot
z_{2}=u_{2}\cdot z_{2}^{\prime }$ there is some good $a_{2}$ such that $%
z_{2a_{2}}>z_{2a_{2}}^{\prime }$. Continuing in this fashion we build a
sequence $1,a_{1},2,a_{2},3,a_{3},\cdots $, such that $%
\{z_{ka_{k-1}}<z_{ka_{k-1}}^{\prime }$ and $z_{ka_{k}}>z_{ka_{k}}^{\prime
}\} $ for all $k\geq 2$. This sequence must cycle, i. e., we must reach $%
K,a_{K}$ such that $z_{Ka_{K}}>z_{Ka_{K}}^{\prime }$ and $z_{\widetilde{k}%
a_{K}}<z_{\widetilde{k}a_{K}}^{\prime }$ for some $\widetilde{k},1\leq 
\widetilde{k}\leq K-1$. Without loss we label $\widetilde{k}$ as $1$, and
the corresponding cycle as $\mathcal{C}$.

From the reasoning above it follows that for $z^{\prime \prime }=\frac{%
(z+z\prime )}{2}$, an efficient allocation, there is a cycle in $\Gamma
(z^{\prime \prime })$. Then the argument in Section $a)$. implies that $u$
is not in $\mathcal{U}^{\ast }(N,A)$, as was to be proved.

\subsection{Lemma 2}

\textit{Case of goods}\textbf{. }Fix $\mathcal{Q},U,z$ as in statement $i)$
and~(\ref{42}). We set $p_{a}=\frac{u_{ia}}{U_{i}}$ for all $i$ such that $%
z_{ia}>0$ and note that $p\cdot z_{i}=1$ for all $i$. For all $a$ such that $%
z_{ia}=0$ we have $\frac{u_{ia}}{U_{i}}\leq p_{a}$: therefore $z_{i}$ is
agent $i$'s Walrasian demand at price $p$, and $z$ is a competitive
allocation.

\noindent Conversely let $z\in \Phi (N,A)$ and $p$ meet~(\ref{27}). Recall $%
U\gg 0$ because each agent $i$ likes at least one good. If $p_{a}=0$ then
nobody likes good $a$ (if $u_{ia}>0$ then $i$'s demand is infinite, a
contradiction of~(\ref{27})) and system~(\ref{42}) holds for $a$. Consider
now the support $A^{\ast }$ of $p_{a}$. Because $z_{i}$ is $i$'s demand at
price $p$ the ratio $\frac{u_{ia}}{p_{a}}$ is a constant $\pi _{i}$ over the
support of $z_{i}$, and we have: $\sum_{A^{\ast }}u_{ia}z_{ia}=\pi
_{i}(\sum_{A^{\ast }}p_{a}z_{ia})\Longrightarrow \pi _{i}=U_{i}$. So $\frac{%
u_{ia}}{U_{i}}=p_{a}$ whenever $i$ consumes $a$ and $\frac{u_{ib}}{U_{i}}%
\leq p_{b}$ if $i$ does not eat any $b$, as required by system~(\ref{42}).
Note that this argument implies that the competitive price $p$ is unique.

\noindent \textit{Case of bads}. Fix $\mathcal{Q},U,z$ meeting~(\ref{43})
and $U\gg 0$. Define $A^{0}=\{a\in A|u_{ia}=0$ for some $i\in N\}$ and set $%
p_{a}=0$ for those bads. By~(\ref{43}) the bads in $A^{0}$ can only be eaten
by agents who don't mind them: $z_{ia}>0\Longrightarrow u_{ia}=0$. Next in
the restriction of $\mathcal{Q}$\ to $A\diagdown A^{0}$ the same utility
profile $U$ is still feasible, strictly positive, and meets~(\ref{43}). Like
in the above argument we set $p_{a}=\frac{u_{ia}}{U_{i}}$ for all $i$ who
eat some $a$, and check that we have constructed a competitive price in the
sense of Definition 4.

For the converse statement, recall that~(\ref{41}),~(\ref{49}) together
imply $U\gg 0$, and mimic the argument in the case of goods.

\subsection{Proposition 2}

\textbf{Statement} $i)$ the $F^{c}$ for goods is Resource Monotonic.

\noindent We first generalize the definition of $F^{c},f^{c}$ to problems
where the endowment $\omega _{a}$ of each good is arbitrary, and let the
reader check that the system~(\ref{42}) capturing the optimal allocations $%
f^{c}(N,A,\omega ,u)$ is unchanged. Then we fix $N,A,u,\omega ,\omega
^{\prime }$ such that $\omega \leq \omega ^{\prime }$. We assume without
loss of generality that $u$ contains no null row or column (all agents are
interested and all goods are useful).For $\lambda \in \lbrack 0,1]$ we write 
$\omega ^{\lambda }=(1-\lambda )\omega +\lambda \omega ^{\prime }$, and for
every forest $\Gamma $ in $N\times A$ we define%
\begin{equation*}
\mathcal{B}(\Gamma )=\{\lambda \in \lbrack 0,1]|\exists z\in
f^{c}(N,A,\omega ^{\lambda },u):\Gamma (z)=\Gamma \}
\end{equation*}%
Note that $\mathcal{B}(\Gamma )$ can be empty or a singleton, but if it is
not, then it is an interval. To see this take $z\in f^{c}(\omega ^{\lambda
}),z^{\prime }\in f^{c}(\omega ^{\lambda ^{\prime }})$ such that $\Gamma
(z)=\Gamma (z^{\prime })$. For any $\omega ^{\prime \prime }=(1-\mu )\omega
^{\lambda }+\mu \omega ^{\lambda ^{\prime }}$ the allocation $z^{\prime
\prime }=(1-\mu )z+\mu z^{\prime }$ is feasible, $z^{^{\prime \prime }}\in
\Phi (N,A,\omega ^{\prime \prime })$, the forest $\Gamma (z^{\prime \prime
}) $ is unchanged, and the system~(\ref{43}), which holds at $z$ and $%
z^{\prime }$, also holds at $z^{\prime \prime }$. Thus $z^{\prime \prime
}\in f^{c}(\omega ^{\prime \prime })$ and the claim is proven.

Next we check that inside an interval $\mathcal{B}(\Gamma )$ the rule $F^{c}$
is resource monotonic. The forest $\Gamma $ is a union of trees. If a tree
contains a single agent $i$, she eats (in full) the same subset of goods for
any $\lambda $ in $\mathcal{B}(\Gamma )$, hence her utility increases weakly
in $\lambda $. If a sub-tree of $\Gamma $ connects the subset $S$ of agents,
then system~(\ref{43}) fixes the direction of the utility profile $%
(U_{i})_{i\in S}$, because along a path of $\Gamma $ the equalities $\frac{%
u_{ia}}{U_{i}}=\frac{u_{ja}}{U_{j}}$ ensure that all ratios $\frac{U_{i}}{%
U_{j}}$ are independent of $\lambda $ in $\mathcal{B}(\Gamma )$. As $\lambda 
$ increases in $\mathcal{B}(\Gamma )$ the agents in $S$ together eat the
same subset of goods, therefore the $U_{i}$-s increase weakly by efficiency.

Finally Lemma 1 implies that the finite set of intervals $\mathcal{B}(\Gamma
)$ cover $[0,1]$. On each true interval (not a singleton) the utility
profile $U^{\lambda }=F(N,A,\omega ^{\lambda },u)$ and there is at most a
finite set of isolated points not contained in any true interval. Moreover
the mapping $\lambda \rightarrow U^{\lambda }$ is continuous because $\omega
\rightarrow U(\omega )$ is (an easy consequence of Berge Theorem). The
desired conclusion $U(\omega )\leq U(\omega ^{\prime })$ follows.\smallskip

\noindent \textit{Remark: Sziklai and Segal-Halevi (\cite{SS1}) prove that
the Competitive solution is Resource Monotonic in the general cake-division
problem, which implies statement }$i)$ \textit{in Theorem 2. They show that
as the cake increases the (normalized) price of the old cake goes down, a
different proof technique than ours.\smallskip }

\textbf{Statement }$ii)$ First we repeat the argument in Subsection 7.2 of 
\cite{BMSY}, focusing on a simple two-person, two-bad example. Suppose the
efficient rule $F$ meets RM and FSG and consider the problem%
\begin{equation*}
\mathcal{Q}=%
\begin{array}{ccc}
\text{bads} & a & b \\ 
u_{1} & 1 & 4 \\ 
u_{2} & 4 & 1%
\end{array}%
\end{equation*}%
Set $U=F(\mathcal{Q})$. Because $(1,1)$ is an efficient disutility profile
and $F$ is efficient, one $U_{i}$ is bounded above by $1$, say $U_{1}\leq 1$%
. Then we define%
\begin{equation*}
\mathcal{Q}^{\prime }=%
\begin{array}{ccc}
\text{bads} & \frac{1}{9}a & b \\ 
u_{1} & 1/9 & 4 \\ 
u_{2} & 4/9 & 1%
\end{array}%
\end{equation*}%
(where we treat $\frac{1}{9}a$ as a whole bad) and pick $z^{\prime }\in f(%
\mathcal{Q}^{\prime })$. By FSG and feasibility:%
\begin{equation*}
z_{2b}^{\prime }\leq u_{2}\cdot z_{2}^{\prime }\leq u_{2}\cdot (\frac{1}{2}%
e^{A^{\prime }})=\frac{13}{18}
\end{equation*}%
\begin{equation*}
\Longrightarrow z_{1b}^{\prime }\geq \frac{5}{18}\Longrightarrow u_{1}\cdot
z_{1}^{\prime }=U_{1}^{\prime }\geq \frac{10}{9}>U_{1}
\end{equation*}%
contradicting RM.

We generalize the example , first to the case where $n=2n^{\prime }$ is
even, $n^{\prime }\geq 2$. Fix two bads $a,b$. At $\mathcal{Q}$ we have $%
n^{\prime }$ agents with $u_{i}=(1,5n^{\prime }),$ $i\in N_{1},$ and $%
n^{\prime }$ agents with $u_{j}=(5n^{\prime },1),$ $j\in N_{2}$. The profile 
$U=\frac{1}{n^{\prime }}e^{N}$ is feasible. Also, at an efficient profile if
at least one in $N_{1}$ eats some $b$, then no one in $N_{2}$ eats any $a$,
and vice versa. Thus at $U=F(\mathcal{Q})$ at least one of $U_{N_{1}}\leq 1$
or $U_{N_{2}}\leq 1$ is true, say $U_{N_{1}}\leq 1$. Then define $\mathcal{Q}%
^{\prime }=(\frac{1}{10n^{\prime }}a,b)$ and use again FSG and feasibility:%
\begin{equation*}
\text{for }j\in N_{2}\text{: }z_{jb}^{\prime }\leq u_{j}\cdot z_{j}^{\prime
}\leq u_{j}\cdot (\frac{1}{2n^{\prime }}e^{A^{\prime }})=\frac{3}{4n^{\prime
}}
\end{equation*}%
\begin{equation*}
\Longrightarrow \sum_{i\in N_{1}}z_{ib}^{\prime }\geq \frac{1}{4}%
\Longrightarrow U_{N_{1}}^{\prime }\geq \frac{5}{4}n^{\prime }>U_{N_{1}}
\end{equation*}%
This contradicts RM. The case $n=2n^{\prime }+1$ odd is very similar, except
that the two groups are of size $n^{\prime }$ and $n^{\prime }+1$, with the
same utilities as above. If we have more than two bads, say $c,d,\ldots $,
we assume their disutilities are very small with respect those for $a,b$.

\subsection{Lemma 4}

We give the proof in the case of bads, because the definition of the
egalitarian allocation is simpler in this case. We omit the slightly more
involved argument for goods, where we must deal with the leximin ordering
(Definition 3).

We fix $\mathcal{Q}$, and $u_{i}^{\prime }$ as in the statement of the
Lemma, in particular $\Gamma (z^{\prime })=\Gamma (z)$. Set%
\begin{equation*}
\lambda =\frac{u_{j}\cdot z_{j}}{u_{j}\cdot e^{A}}\text{ for all }j\text{ ; }%
\lambda ^{\prime }=\frac{u_{i}^{\prime }\cdot z_{i}^{\prime }}{u_{i}^{\prime
}\cdot e^{A}}=\frac{u_{j}\cdot z_{j}^{\prime }}{u_{j}\cdot e^{A}}\text{ for
all }j\neq i
\end{equation*}%
Inequalities~(\ref{22}) (reversed for bads!) imply%
\begin{equation*}
\mu =\frac{u_{i}^{\prime }\cdot z_{i}}{u_{i}^{\prime }\cdot e^{A}}>\frac{%
u_{i}\cdot z_{i}}{u_{i}\cdot e^{A}}=\lambda
\end{equation*}%
because the ratio increases when we first increase $u_{ia}$ to $%
u_{ia}^{\prime }$ on $i$'s winning bids (the rest of the numerator is not
larger than the rest of the denominator), then again when we decrease $%
u_{ia} $ to $u_{ia}^{\prime }$ on $i$'s losing bids. Next we assume $\lambda
^{\prime }\geq \mu $ and derive a contradiction. It implies%
\begin{equation*}
u_{j}\cdot z_{j}^{\prime }\geq (u_{j}\cdot e^{A})\mu >(u_{j}\cdot
e^{A})\lambda =u_{j}\cdot z_{j}\text{ for all }j\neq i
\end{equation*}%
and $u_{i}^{\prime }\cdot z_{i}^{\prime }\geq u_{i}^{\prime }\cdot z_{i}$;
together these inequalities contradict the efficiency of $z^{\prime }$ at $%
(u_{i}^{\prime },u_{-i})$. Therefore $\lambda ^{\prime }<\mu $ and $%
u_{i}^{\prime }\cdot z_{i}^{\prime }<u_{i}^{\prime }\cdot z_{i}$. Finally $%
\Gamma (z^{\prime })=\Gamma (z)$ implies that a winning (resp. losing) bid
at $z$ remains winning (resp. losing) at $z^{\prime }$, so that $%
u_{i}^{\prime }\cdot z_{i}^{\prime }-u_{i}\cdot z_{i}^{\prime
}=u_{i}^{\prime }\cdot z_{i}-u_{i}\cdot z_{i}$ and $u_{i}\cdot z_{i}^{\prime
}<u_{i}\cdot z_{i}$ follows as desired.

\subsection{Proposition 3}

\noindent We already checked that the rule $f^{c}$\ meets ILB; also ETE and
FSG are clear. Conversely we fix $f$ meeting EFF, ETE or FSG, and ILB and an
arbitrary problem $\mathcal{Q}=(N,A,u)$, where $A$ contains goods or bads.
In the proof we consider several problems $(N,A,v)$ where $v$ varies in $%
\mathbb{R}
_{+}^{N\times A}$, and for simplicity we write $f(v)$ in lieu of $f(N,A,v)$.

We pick $z\in f^{c}(u\mathcal{)}$ and check that $z\in f(u\mathcal{)}$ as
well. Set $U_{i}=u_{i}\cdot z_{i}$ and let $p$ be the competitive price at $%
z $. In the proof of Lemma 2 we saw that $p_{a}=\frac{u_{ia}}{U_{i}}$ for
all $i$ such that $z_{ia}>0$, and for all $j$ we have $p_{a}\geq \frac{u_{ja}%
}{U_{j}}$ (resp. $p_{a}\leq \frac{u_{ja}}{U_{j}}$) if we divide goods (resp.
bads). Moreover $p\cdot z_{i}=1$ for all $i$, and $p\cdot e^{A}=n$.

Consider the problem $\mathcal{Q}^{\ast }=(N,A,w)$ where $w_{i}=p$ for all $%
i $. The equal split allocation is efficient in $\mathcal{Q}^{\ast }$
therefore ETE implies $F(w)=e^{N}$ and so does FSG, because $p\cdot (\frac{1%
}{n}e^{A})=1$. Now if we set $\widetilde{w}_{i}=U_{i}p$ the scale invariance
property of $F$ (Definition 2) gives $F(\widetilde{w})=U$; moreover $z\in f(%
\widetilde{w})$ because $\widetilde{w}_{i}\cdot z=U_{i}$ for all $i$. If $%
z_{ia}>0$ we have $u_{ia}=U_{i}p_{a}=\widetilde{w}_{ia}$; if $z_{ia}=0$ we
have similarly $u_{ia}\leq \widetilde{w}_{ia}$ for the goods case, or $%
u_{ia}\geq \widetilde{w}_{ia}$ if bads. Apply finally ILB: after lowering
(for goods) or raising (for bads) every lost bid $\widetilde{w}_{ia}$ to $%
u_{ia}$, the allocation $z$ is still in $f(u)$, as desired. $\blacksquare
\smallskip $

Finally we check that ILB is in fact a weaker form of Maskin Monotonicity
(MM). We do this in the case of bads only, as both cases are similar.
Individual allocations $z_{i}$ vary in $[0,1]^{A}$\ and utilities in $%
\mathbb{R}
_{+}^{A}$, so the MM axiom for rule $f$\ means that for any two problems $%
\mathcal{Q},\mathcal{Q}^{\prime }$ on $N,A$\ and $z\in f(\mathcal{Q})$ we
have%
\begin{equation}
\{\forall i\in N,\forall w\in \lbrack 0,1]^{A}\text{ \ }u_{i}\cdot z_{i}\leq
u_{i}\cdot w\Longrightarrow u_{i}^{\prime }\cdot z_{i}\leq u_{i}^{\prime
}\cdot w\}\Longrightarrow z\in f(\mathcal{Q}^{\prime })  \label{45}
\end{equation}%
We fix $\mathcal{Q},i\in N$ and $z\in f(\mathcal{Q})$, and for $\varepsilon
=0,1$\ we write $A^{\varepsilon }=\{a|z_{ia}=\varepsilon \}$\ and $%
A^{+}=A\diagdown (A^{0}\cup A^{1})$. The implication in the premises of (\ref%
{45}) reads%
\begin{equation*}
\forall w\in \lbrack 0,1]^{A}\text{ }u_{i}\cdot (w-z_{i})\geq
0\Longrightarrow u_{i}^{\prime }\cdot (w-z_{i})\geq 0
\end{equation*}%
The cone generated by the vectors $w-z_{i}$ when $w$ covers $[0,1]^{A}$ is $%
C=\{\delta \in 
\mathbb{R}
^{A}|\delta _{a}\geq 0$ for $a\in A^{0}$, $\delta _{a}\leq 0$ for $a\in
A^{1}\}$. By Farkas Lemma the implication $\{\forall \delta \in C:$ $%
u_{i}\cdot \delta \geq 0\Longrightarrow u_{i}^{\prime }\cdot \delta \geq 0\}$
means that, up to a scaling factor,%
\begin{equation*}
u_{ia}^{\prime }=u_{ia}\text{ on }A^{+}\text{ ; }u_{ia}^{\prime }\geq u_{ia}%
\text{ on }A^{0}\text{ ; }u_{ia}^{\prime }\leq u_{ia}\text{ on }A^{1}
\end{equation*}%
Thus MM says that after lowering a lost bid, or increasing a winning one,
the initial allocation will remain in the selected set. Now ILB only
considers raising a lost bid, so it is only \textquotedblleft
half\textquotedblright\ of MM. The Competitive rule does not meet the other
half of MM.

\subsection{Theorem 1}

\paragraph{Statement $i)$}

Fix $\mathcal{Q}$ and recall from Lemma 1 that each $U\in F^{c}(\mathcal{Q})$
is strictly positive and achieved some $z\in f^{c}(\mathcal{Q})$ such that $%
\Gamma (z)$\ is a forest. There are finitely many (bipartite) forests in $%
N\times A$ therefore it is enough to check that to each forest $\Gamma $
corresponds at most one $U$ in $F^{c}(\mathcal{Q})$. Consider a tree $T$ in $%
\Gamma $ with vertices $N_{0},A_{0}$. If agents $i,j\in N_{0}$ are both
linked to $a\in A_{0}$, system~(\ref{43}) implies that $U_{i},U_{j}$ are
proportional to $u_{ia},u_{ja}$ . Repeating this observation along the paths
of $T$ we see that the profile $(U_{i})_{i\in N_{0}}$ is determined up to a
multiplicative constant. Now in total the agents in $N_{0}$ consume exactly $%
A_{0}$ so by efficiency we cannot have two distinct $(U_{i})_{i\in N_{0}}$
meeting~(\ref{43}).\footnote{%
Note that the finiteness result holds even if we drop requirement~(\ref{49})
in Definition 3 but still insist that a competitive allocation be efficient.
If $A^{0}$ is the set of bads $a$ such that $u_{ia}=0$ for some $i$, then
some items in $A^{0}$ can have a positive price, and be eaten by agents who
do not mind them, eat only in $A^{0}$, and enjoy a disutility of zero; while
the other bads in $A^{0}$ have zero price, are also eaten by agents who do
not mind them but those agents eat also some real bads in $A\diagdown A^{0}$%
. For each such partition of $A^{0}$ there are finitely many competitive
disutility profiles.}

\paragraph{Statement $ii)$}

\noindent Case 1: $n>m$. We adapt the canonical example before Theorem 1 as
follows. For agent $i,1\leq i\leq m,$ set as before $%
u_{ia_{i}}=1,u_{ia_{j}}=3$ for $j\neq i$, and for agents $m+1$ to $n$ pick $%
u_{ia}=1$ for all $a$. Then for any $q,1\leq q\leq n-1$, the allocation%
\begin{equation*}
z_{i}=\frac{m}{n}e^{a_{i}}\text{ for }1\leq i\leq q\text{ ; }z_{j}=e^{a_{j}}%
\text{ for }q+1\leq j\leq m
\end{equation*}%
\begin{equation*}
z_{j}=\frac{1}{n}e^{\{a_{1},\cdots ,a_{q}\}}\text{ for }m+1\leq j\leq n
\end{equation*}%
generalizing~(\ref{46}), is still competitive.

\noindent Case 2: $m=n+1$. We use the following example%
\begin{equation*}
\begin{array}{cccccc}
\text{bad} & a_{1} & \cdots & \cdots & a_{n} & a_{n+1} \\ 
u_{1} & 1 & 3 & 3 & 3 & 1 \\ 
\cdots & 3 & 1 & 3 & 3 & 1 \\ 
\cdots & 3 & 3 & 1 & 3 & 1 \\ 
u_{n} & 3 & 3 & 3 & 1 & 1%
\end{array}%
\end{equation*}

\noindent For any subset of agents $N^{\ast }\subseteq N$ the allocation
where those agents share equally the bad $n+1$, while bad $a_{i},1\leq i\leq
n$ goes to agent $i$, is competitive with prices $p_{a_{n+1}}=p_{a_{i}}=%
\frac{n^{\ast }}{n^{\ast }+1}$ for $i\in N^{\ast }$, $p_{a_{j}}=1$ for $j\in
N\diagdown N^{\ast }$.

For the case $m>n$ we take a disutility matrix with $m-n$ copies of the last
column. We omit the details as well as the easy argument for the case $n=m$.

\paragraph{Statement $iii)$ and oddness of $|F^{c}(\mathcal{Q})|$}

We fix $\mathcal{Q=}(\{1,2\},A,u)$. We label the bads $k\in \{1,\cdots ,m\}$
so that the ratios $\frac{u_{1k}}{u_{2k}}$ increase weakly in $k$, with the
convention $\frac{1}{0}=\infty $.

\noindent \textit{Step 1}. Suppose $\frac{u_{1k}}{u_{2k}}=\frac{u_{1(k+1)}}{%
u_{2(k+1)}}$. If $p$ is a competitive price we have%
\begin{equation}
\frac{p_{k}}{p_{k+1}}=\frac{u_{ik}}{u_{i(k+1)}}\text{ for }i=1,2  \label{4}
\end{equation}%
Indeed if one of $i=1,2$ eats both $k$ and $k+1$,~(\ref{4}) follows by the
linearity of preferences, If on the contrary $i$ eats bad $k$ and $j$ eats
bad $k+1$, then~(\ref{43}) gives $\frac{u_{ik}}{p_{k}}\geq \frac{u_{i(k+1)}}{%
p_{k+1}}$ and $\frac{u_{j(k+1)}}{p_{k+1}}\geq \frac{u_{jk}}{p_{k}}$.

So for a given amount of money spent by $i$ on bads $k$ and $k+1$, she gets
the same disutility no matter how she splits this expense between the two
bads. Hence an interval of competitive allocations obtains by shifting the
consumption of $k$ and $k+1$ while keeping the total expense on these two
bads fixed for each agent. They all give the same disutility profile and use
the same price. So if we merge $k$ and $k+1$ into a bad $k^{\ast }$ with
disutilities $u_{ik^{\ast }}=u_{ik}+u_{i(k+1)}$, all the allocations $z\in
f^{c}(\mathcal{Q)}$ become a single competitive allocation for the new price 
$p_{k^{\ast }}=p_{k}+p_{k+1}$, with $p$ unchanged elsewhere. When we
successively merge all the bads sharing the same ratio $\frac{u_{1k}}{u_{2k}}
$, the number $|F^{c}(\mathcal{Q)}|$ does not change, and we reach a problem
with fewer bads where the ratios $\frac{u_{1k}}{u_{2k}}$ increase strictly
in $k$. So we only need to prove the statement in this case.\smallskip

\noindent \textit{Step 2. }Efficiency means that if $1$ eats some $k$ and $2$
some $k^{\prime }$, then $k\leq k^{\prime }$. In particular the agents split
at most one bad, and efficient allocations $z$ are of two types. In a $k/k+1$%
\textit{-cut }$1$ eats all bads $\ell \in \{1,\cdots ,k\}$ while $2$ eats
all $\ell \in \{k+1,\cdots ,m\}$; in a $k$\textit{-split }$1$ and $2$ share
bad $k$, while bads $\ell \in \{1,\cdots ,k-1\}$ go to $1$, and $\ell \in
\{k+1,\cdots ,m\}$ to $2$.

By Definition 4 if the $k/k+1$\textit{-cut} is in $f^{c}(\mathcal{Q)}$, its
(normalized) price is%
\begin{equation*}
p_{\ell }=\frac{u_{1\ell }}{U_{1}(k)}\text{ for }\ell \leq k\text{ ; }%
p_{\ell }=\frac{u_{2\ell }}{U_{2}(k+1)}\text{ for }\ell \geq k+1
\end{equation*}%
where we use the notation $U_{1}(k)=\sum_{1}^{k}u_{1\ell }$, $%
U_{2}(k)=\sum_{k}^{m}u_{2\ell }$. System~(\ref{43}) has two parts: agent $1$%
's demand at this price is $e^{\{1,\cdots ,k\}}$:%
\begin{equation*}
\{\frac{u_{1\ell }}{p_{\ell }}\leq \frac{u_{1\ell ^{\prime }}}{p_{\ell
^{\prime }}}\text{ for all }\ell \leq k<k+1\leq \ell ^{\prime
}\}\Longleftrightarrow \frac{U_{1}(k)}{U_{2}(k+1)}\leq \frac{u_{1(k+1)}}{%
u_{2(k+1)}}
\end{equation*}%
and agent $2$'s demand is $e^{\{k+1,\cdots ,m\}}$:%
\begin{equation*}
\{\frac{u_{2\ell ^{\prime }}}{p_{\ell ^{\prime }}}\leq \frac{u_{2\ell }}{%
p_{\ell }}\text{ for all }\ell \leq k<k+1\leq \ell ^{\prime
}\}\Longleftrightarrow \frac{u_{1k}}{u_{2k}}\leq \frac{U_{1}(k)}{U_{2}(k+1)}
\end{equation*}%
Thus the $k/k+1$\textit{-}cut is in $f^{c}(\mathcal{Q)}$ if and only%
\begin{equation}
\frac{u_{1k}}{u_{2k}}\leq \frac{U_{1}(k)}{U_{2}(k+1)}\leq \frac{u_{1(k+1)}}{%
u_{2(k+1)}}  \label{5}
\end{equation}

We turn to a $k$\textit{-split }allocation. If it is competitive, by
Definition 4 the corresponding \textit{non normalized} price is%
\begin{equation*}
p_{\ell }=\frac{u_{1\ell }}{u_{1k}}\text{ for }\ell \leq k-1\text{ ; }p_{k}=1%
\text{ ; }p_{\ell }=\frac{u_{2\ell }}{u_{2k}}\text{ for }\ell \geq k+1
\end{equation*}%
and the inequalities $\frac{u_{1\ell }}{p_{\ell }}\leq \frac{u_{1\ell
^{\prime }}}{p_{\ell ^{\prime }}}$ and $\frac{u_{2\ell ^{\prime }}}{p_{\ell
^{\prime }}}\leq \frac{u_{2\ell }}{p_{\ell }}$ hold by construction for $%
\ell \leq k\leq \ell ^{\prime }$. To reach a competitive allocation it
suffices to find a split of bad $k$ for which $z_{1}$ and $z_{2}$ have the
same cost:%
\begin{equation}
{\Large \{}\sum_{1}^{k-1}p_{\ell }+x=1-x+\sum_{k+1}^{m}p_{\ell ^{\prime }}%
\text{ for some }x\in ]0,1[{\Large \}\Longleftrightarrow |}\frac{U_{1}(k-1)}{%
u_{1k}}-\frac{U_{2}(k+1)}{u_{2k}}|<1  \label{6}
\end{equation}%
Note that for $k=1$ inequalities~(\ref{6}) reduce to $U_{2}(2)<u_{21}$, and
for $k=m$ to $U_{1}(m-1)<u_{1m}$.

We see that for each $k$ there is at most one $k$-split allocation in $f^{c}(%
\mathcal{Q)}$. So $|F^{c}(\mathcal{Q)}|$ is at most $2m-1$, because the
maximal number of $k/k+1$\textit{-}cuts and $k$-split allocations is
respectively $m-1$ and $m$.\smallskip

\noindent \textit{Step 3. }We use the notation $(x)_{+}=\max \{x,0\}$ to
give an example where this bound is achieved:%
\begin{equation*}
u_{1k}=2^{(k-2)_{+}}\text{ for }1\leq k\leq m-1\text{ ; }u_{1m}=2^{m-2}+1
\end{equation*}%
\begin{equation*}
u_{21}=2^{m-2}+1\text{ ; }u_{2k}=2^{(m-1-k)_{+}}\text{ for }2\leq k\leq m%
\text{ }
\end{equation*}%
Check first $U_{1}(k-1)=u_{1k}$ and $U_{2}(k+1)=u_{2k}$ for $2\leq k\leq m-1$%
; also $U_{2}(2)=U_{1}(m-1)=2^{m-2}<u_{21}=u_{1m}$ so~(\ref{6}) holds for
all $k$. Next $\frac{U_{1}(k)}{U_{2}(k+1)}=\frac{u_{1(k+1)}}{u_{2k}}$ for $%
2\leq k\leq m-2$, so that~(\ref{5}) is clear for such $k$. And~(\ref{5})
holds as well for $k=1,m-1$.

This example is \ clearly robust: small perturbations of the disutility
matrix preserve $|F^{c}(\mathcal{Q)}|$.\smallskip

\noindent \textit{Step 4. }Oddness of $|F^{c}(\mathcal{Q})|$. Recall that $%
F^{c}(\mathcal{Q})$ is the set of critical points of the Nash product in%
\textit{\ }$\Psi ^{eff}(\mathcal{Q})$.The $k/k+1$-cut allocations are the
extreme points of the set of feasible allocations $\Phi (N,A)$, and their
utility profiles are the extreme points\footnote{%
An $U\in \Psi ^{eff}(\mathcal{Q})$ is \textit{extreme }if it is not between
two other points of $\Psi ^{eff}(\mathcal{Q})$.} of $\Psi ^{eff}(\mathcal{Q}%
) $ if the ratios $\frac{u_{1k}}{u_{2k}}$ increase strictly in $k$.
Excluding the set of utility profiles $u$ such that ${\Large |}\frac{%
U_{1}(k-1)}{u_{1k}}-\frac{U_{2}(k+1)}{u_{2k}}|=1$ (see~(\ref{6})), it
follows that the $k/k+1$ cut is competitive if and only if it is a local
minimum of $\mathcal{N}$. On the other hand the utility profile of a $k$%
-split allocation is on a one-dimensional face of $\Psi ^{eff}(\mathcal{Q})$%
, and is competitive if and only if it is a local maximum of $\mathcal{N}$.
Then the statement follows from the fact that if a continuous non-negative
function on the interval is zero at the end-points, the number of its local
maxima exceeds the number of its local minima (different than the
end-points) by one: the extrema alternate and the closest to the end-points
are the maxima.

Note that the above argument implies that in a typical problem with two
agents, if $|F^{c}(\mathcal{Q})|=1$ then the competitive allocation is a $%
k/k+1$-split, and if $|F^{c}(\mathcal{Q})|\geq 2$, at least one $k$-cut
allocation is competitive.

\paragraph{Statement $iv)$ and oddness of $|F^{c}(\mathcal{Q})|$}

We fix $\mathcal{Q=}(N,\{a,b\},u)$ and label the agents $i\in \{1,\cdots
,n\} $ in such a way that the ratios $\frac{u_{ia}}{u_{ib}}$ increase weakly
in $i $.

\noindent \textit{Step 1. Assume that the sequence }$\frac{u_{ia}}{u_{ib}}$%
\textit{\ increases strictly}. If $z$ is an efficient allocation, then for
all $i,j$ $\{z_{ia}>0$ and $z_{jb}>0\}$ implies $i\leq j$. In particular at
most one agent is eating both goods, and we have two types of efficient and
envy-free allocations. The $i/i+1$\textit{-cut} $z^{i/i+1}$ is defined for $%
1\leq i\leq n-1$ by: $z_{j}^{i/i+1}=(\frac{1}{i},0)$ for $j\leq i$, and $%
z_{j}^{i/i+1}=(0,\frac{1}{n-i})$ for $j\geq i+1$. Next for $2\leq i\leq n-1$
the allocation $z$ is an $i$\textit{-split} if there are numbers $x,y\ $such
that%
\begin{equation}
z_{j}=(\frac{1-x}{i-1},0)\text{ for }j\leq i-1\text{ ; }z_{j}=(0,\frac{1-y}{%
n-i})\text{ for }j\geq i+1  \label{8}
\end{equation}%
\begin{equation}
z_{i}=(x,y)\text{ with }0\leq x\leq \frac{1}{i}\text{, }0\leq y\leq \frac{1}{%
n-i+1}  \label{11}
\end{equation}%
Also, $z$ is a $1$-split if $z_{1}=(1,y)$ and $z_{j}=(0,\frac{1-y}{n-1})$
for $j\geq 2$; and $z$ is a $n$-split if $z_{n}=(x,1)$ and $z_{j}=(\frac{1-x%
}{n-1},0)$ for $j\leq n-1$. Note that the cut $z^{i/i+1}$ is both an $i$%
-split and an $i+1$-split.

If the sequence $\frac{u_{ia}}{u_{ib}}$ increases strictly, it is clear that
an efficient and envy-free allocation must be an $i$-split. In the next Step
we show that this is still true, welfare-wise, if that sequence increases
only weakly, then we provide a full characterization in Step 3.\smallskip

\noindent \textit{Step 2}. Assume the sequence $\frac{u_{ia}}{u_{ib}}$
increases only weakly, for instance $\frac{u_{ia}}{u_{ib}}=\frac{u_{(i+1)a}}{%
u_{(i+1)b}}$. Then if $z$ is efficient and envy-free we may have $%
z_{(i+1)a}>0$ and $z_{ib}>0$, however we can find $z^{\prime }$ delivering
the same disutility profile and such that one of $z_{(i+1)a}^{\prime }$ and $%
z_{ib}^{\prime }$ is zero. Indeed No Envy and the fact that $u_{i}$ and $%
u_{i+1}$ are parallel gives $u_{i}\cdot z_{i}=u_{i}\cdot z_{i+1}$ and $%
u_{i+1}\cdot z_{i+1}=u_{i+1}\cdot z_{i}$, from which the claim follows
easily. We conclude that the $i$\textit{-}split allocations contain,
utility-wise, all efficient and envy-free allocations.\smallskip

\noindent \textit{Step 3.} If the cut $z^{i/i+1}$ is in $f^{c}(\mathcal{Q})$%
, the corresponding price is $p=(i,n-i)$, and the system~(\ref{43}) reads $%
\frac{u_{ja}}{i}\leq \frac{u_{jb}}{n-i}$ for $j\leq i$, $\frac{u_{jb}}{n-i}%
\leq \frac{u_{ja}}{i}$ for $j\geq i+1$, which boils down to%
\begin{equation}
\frac{u_{ia}}{u_{ib}}\leq \frac{i}{n-i}\leq \frac{u_{(i+1)a}}{u_{(i+1)b}}%
\text{ for }1\leq i\leq n-1  \label{9}
\end{equation}

For $2\leq i\leq n-1$ if the $i$\textit{-}split allocation $z$~(\ref{8}) is
in $f^{c}(\mathcal{Q})$, the (normalized) price must be $p=n(\frac{u_{ia}}{%
u_{ia}+u_{ib}},\frac{u_{ib}}{u_{ia}+u_{ib}})$ and each agent must be
spending exactly $1$:%
\begin{equation*}
p_{a}\frac{1-x}{i-1}=p_{b}\frac{1-y}{n-i}=p_{a}x+p_{b}y=1
\end{equation*}%
which gives%
\begin{equation}
x=\frac{1}{nu_{ia}}{\large ((n-i+1)u}_{ia}-(i-1)u_{ib}{\large )}\text{ ; }y=%
\frac{1}{nu_{ib}}{\large (iu}_{ib}-(n-i)u_{ia}{\large )}  \label{50}
\end{equation}%
We let the reader check that these formulas are still valid when $i=1$ or $%
i=n-1$.

An $i$\textit{-}split allocation $z$ is \textit{strict }if it is not a cut,
which happens if and only if both $x,y$ in~(\ref{8}) are strictly positive.
By~(\ref{50}), for any $i\in \{1,\cdots ,n\}$ there is a strict $i$-split
allocation that is competitive if and only if

\begin{equation}
\frac{i-1}{n-i+1}<\frac{u_{ia}}{u_{ib}}<\frac{i}{n-i}  \label{10}
\end{equation}%
(with the convention $\frac{1}{0}=\infty $).\smallskip

\noindent \textit{Step 4}. Counting competitive allocations. There are at
most $n$ competitive (strict) $i$-split allocations, and $n-1$ cuts $%
z^{i/i+1}$, hence the upper bound $2n-1$. An example where the bound is
achieved uses any sequence $\frac{u_{ia}}{u_{ib}}$ meeting~(\ref{10}) for
all $i\in \{1,\cdots ,n\}$, as these inequalities imply~(\ref{9}) for all $%
i\in \{1,\cdots ,n-1\}$.\smallskip

\noindent \textit{Step 5}. Oddness of $|F^{c}(\mathcal{Q})|$. For the
utility profiles such that all the inequalities~(\ref{9}) and~(\ref{10}) are
strict, we draw the two sequences $\frac{u_{ia}}{u_{ib}}$ and $\frac{i}{n-i}$
on the real line. Clearly the left-most and the right-most competitive
allocations must be splits: if there is no competitive $i$-split allocation
for $1\leq i\leq i^{\ast }$ then~(\ref{10}) gives successively $\frac{u_{1a}%
}{u_{1b}}>\frac{1}{n-1}$, then $\frac{u_{2a}}{u_{2b}}>\frac{2}{n-2},\cdots ,%
\frac{u_{i^{\ast }a}}{u_{i^{\ast }b}}>\frac{i^{\ast }}{n-i^{\ast }}$, hence
the $i^{\ast }/i^{\ast }+1$-cut is not competitive. Similarly one checks
that between two adjacent competitive split allocations there is exactly one
competitive cut allocation.

\subsection{Proposition 4}

\noindent \textit{Step 1 the case }$m=2$

\noindent As in the previous proof we fix a problem $(N,\{a,b\},u)$ where
the ratios $r_{i}=\frac{u_{ia}}{u_{ib}}$ increase strictly in $i\in
\{1,\cdots ,n\}$. We write $S^{i}$ for the closed rectangle of $i$-split
allocations~(\ref{8}),~(\ref{11}): we have $S^{i}\cap S^{i+1}=\{z^{i/i+1}\}$
for $i=1,\cdots ,n-1$, and $S^{i}\cap S^{j}=\varnothing $ if $i$ and $j$ are
not adjacent. We saw that envy-free and efficient allocations must be in the
connected union of rectangles $\mathcal{B}={\Large \cup }_{i=1}^{n}S^{i}$.
Writing $\mathcal{EF}$ for the set of envy-free allocations, we describe now
the connected components of $\mathcal{A}=\mathcal{B\cap EF}$. Clearly the
set of corresponding disutility profiles has the same number of connected
components.

We let the reader check that the cut $z^{i/i+1}$ is envy-free (EF) if and
only if it is competitive, i. e. inequalities~(\ref{9}) hold, that we
rewrite as:%
\begin{equation}
r_{i}\leq \frac{i}{n-i}\leq r_{i+1}  \label{38}
\end{equation}%
If $z^{i/i+1}$ is EF then both $S^{i}\cap \mathcal{EF}$ and $S^{i+1}\cap 
\mathcal{EF}$ are in the same component of $\mathcal{A}$ as $z^{i/i+1}$,
because they are convex sets containing $z^{i/i+1}$. If both $z^{i-1/i}$ and 
$z^{i/i+1}$ are EF, so is the interval $[z^{i-1/i},z^{i/i+1}]$; then these
two cuts as well as $S^{i}\cap \mathcal{EF}$ are in the same component of $%
\mathcal{A}$. And if $z^{i/i+1}$ is EF but $z^{i-1/i}$ is not, then the
component of $\mathcal{A}$ containing $z^{i/i+1}$ is disjoint from any
component of $\mathcal{A}$ in ${\Large \cup }_{1}^{i-1}S^{j}$ (if any),
because $S^{i}\cap {\Large \cup }_{1}^{i-1}S^{j}=\{z^{i-1/i}\}$; a
symmetrical statement holds if $z^{i-1/i}$ is EF but $z^{i/i+1}$ is not.

Finally if $S^{i}\cap \mathcal{EF}\neq \varnothing $ while neither $%
z^{i-1/i} $ nor $z^{i/i+1}$ is in $\mathcal{EF}$, the convex set $S^{i}\cap 
\mathcal{EF}$ is a connected component of $\mathcal{A}$ because it is
disjoint from $S^{i-1}\cap \mathcal{EF}$ and $S^{i+1}\cap \mathcal{EF}$, and
all three sets are compact. In this case we speak of an interior component
of $\mathcal{A}$. We claim that $S^{i}$ contains an interior component if
and only if%
\begin{equation}
\frac{i-1}{n-i+1}<r_{i-1}<r_{i}<r_{i+1}<\frac{i}{n-i}  \label{12}
\end{equation}%
where for $i=1$ this reduces to the two right-hand inequalities, and for $%
i=n $ to the two left-hand ones. The claim is proven in the next Step.

Now consider a problem with the following configuration:%
\begin{equation*}
r_{1}<r_{2}<\frac{1}{n-1}<\frac{3}{n-3}<r_{3}<r_{4}<r_{5}<\frac{4}{n-4}<
\end{equation*}%
\begin{equation*}
<\frac{6}{n-6}<r_{6}<r_{7}<r_{8}<\frac{7}{n-7}<\frac{9}{n-9}\cdots
\end{equation*}%
By inequalities~(\ref{38}) we have $z^{i/i+1}\in \mathcal{EF}$ for $i=3q-1,$
and $1\leq q\leq \lfloor \frac{n}{3}\rfloor $, and no two of those cuts are
adjacent so they belong to distinct components. Moreover $S^{i}$ contains an
interior component of $\mathcal{A}$ for $i=3q-2,$ and $1\leq q\leq \lfloor 
\frac{n+2}{3}\rfloor $, and only those. So the total number of components of 
$\mathcal{A}$ is $\lfloor \frac{n}{3}\rfloor +\lfloor \frac{n+2}{3}\rfloor
=\lfloor \frac{2n+1}{3}\rfloor $ as desired.

We let the reader check that we cannot reach a larger number of
components.\smallskip

\noindent \textit{Step 2: }$\{S^{i}$ \textit{contains an interior component}$%
\}\Longleftrightarrow \{$\textit{inequalities~(\ref{12}) hold}$\}$

\noindent Pick $z\in S^{i}$ as in~(\ref{8}),~(\ref{11}) and note first that
for $2\leq i\leq n-1$, the envy-freeness inequalities reduce to just four
inequalities: agents $i-1$ and $i$ do not envy each other, and neither do
agents $i$ and $i+1$ (we omit the straightforward argument). Formally%
\begin{equation}
\frac{1}{r_{i+1}}(\frac{1}{n-i}-\frac{n-i+1}{n-i}y)\leq x\leq \frac{1}{r_{i}}%
(\frac{1}{n-i}-\frac{n-i+1}{n-i}y)  \label{13}
\end{equation}%
\begin{equation}
r_{i-1}(\frac{1}{i-1}-\frac{i}{i-1}x)\leq y\leq r_{i}(\frac{1}{i-1}-\frac{i}{%
i-1}x)  \label{14}
\end{equation}%
In the (non negative) space $(x,y)$ define the lines $\Delta (\lambda )$: $%
y=\lambda (\frac{1}{i-1}-\frac{i}{i-1}x)$ and $\Gamma (\mu )$: $x=\mu (\frac{%
1}{n-i}-\frac{n-i+1}{n-i}y)$. As shown on Figure~3 when $\lambda $ varies $%
\Delta (\lambda )$ pivots around $\delta =(\frac{1}{i},0)$, corresponding to 
$z^{i/i+1}$, and similarly $\Gamma (\mu )$ pivots around $\gamma =(0,\frac{1%
}{n-i+1})$, corresponding to $z^{i-1/i}$. The above inequalities say that $%
(x,y)$ is in the cone $\Delta ^{\ast }$ of points below $\Delta (r_{i})$ and
above $\Delta (r_{i-1})$, and also in the cone $\Gamma ^{\ast }$ below $%
\Gamma (\frac{1}{r_{i}})$ and above $\Gamma (\frac{1}{r_{i}+1})$. Thus $%
\delta \in \Gamma ^{\ast }$ if and only if $z^{i/i+1}$ is EF, and $\gamma
\in \Delta ^{\ast }$ if and only if $z^{i-1/i}$ is EF. If neither of these
is true $\gamma $ is above or below $\Delta ^{\ast }$ on the vertical axis
and $\delta $ is to the left or to the right of $\Gamma ^{\ast }$ the
horizontal axis. But if $\gamma $ is below $\Delta ^{\ast }$ while $\delta $
is right of $\Gamma ^{\ast }$, the two cones do not intersect and $S^{i}\cap 
\mathcal{EF}=\varnothing $; ditto if $\gamma $ is above $\Delta ^{\ast }$
while $\delta $ is left of $\Gamma ^{\ast }$ (see Figures 3A,3B,3C).
Moreover $\gamma $ above $\Delta ^{\ast }$ and $\delta $ right of $\Gamma
^{\ast }$ is impossible as it would imply%
\begin{equation*}
\frac{1}{n-i+1}>\frac{r_{i}}{i-1}\text{ and }\frac{1}{i}>\frac{1}{r_{i}(n-i)}
\end{equation*}%
a contradiction. We conclude that $\{S^{i}\cap \mathcal{EF}\neq \varnothing $
and $z^{i-1/i},z^{i/i+1}\notin \mathcal{EF}\}$ holds if and only if $\gamma $
is below $\Delta ^{\ast }$ and $\delta $ is to the left of $\Gamma ^{\ast }$%
, which is exactly the system~(\ref{12}).

\begin{figure}
\centering
\vskip -0.5 cm
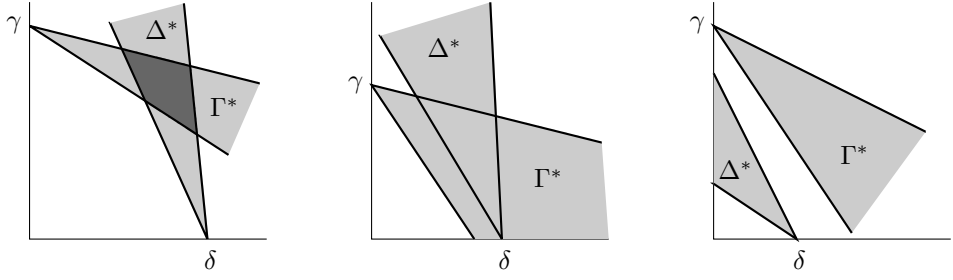
\caption*{Figures 3A, 3B, 3C}
\end{figure}

In the case $i=1$ the EF property of $z$ reduces to~(\ref{13}) and the $i$%
-split allocation has $x=1$. If $r_{1}>\frac{1}{n-1}$ the right-hand
inequality in~(\ref{13}) is impossible with $x=1$, therefore $r_{1}<\frac{1}{%
n-1}$; but then the fact that $z^{1/2}$ is not EF gives (see~(\ref{38})) $%
r_{2}<\frac{1}{n-1}$ as desired. A similar argument applies for the case $%
i=n $.$\smallskip $

\newpage

\noindent \textit{Step 3: general }$m$

\noindent Fix a problem $(N,\{a,b\},u)$ with $\lceil \frac{2n+1}{3}\rceil $
connected components as in Step 1. Given any $m\geq 3$, construct a problem $%
(N,\widetilde{A},\widetilde{u})$ with $\widetilde{A}=\{a,b_{1},\cdots
,b_{m-1}\}$ and for all agents $i$

\begin{equation*}
\widetilde{u}_{ia}=u_{ia}\text{ ; }\widetilde{u}_{ib_{k}}=\frac{1}{m-1}u_{ib}%
\text{ for all }1\leq k\leq m-1
\end{equation*}%
The bads $b_{k}$ are smaller size clones of $b$. If some $\widetilde{z}$ is
efficient and EF in the new problem, then the following allocation $z$ is
efficient and EF in the initial problem:%
\begin{equation*}
z_{ib}=\sum_{1}^{m-1}\widetilde{z}_{ib_{k}}\text{ ; }z_{ia}=\widetilde{z}%
_{ia}
\end{equation*}%
and $z,\widetilde{z}$ deliver the same disutility profile. Therefore in the
two problems the sets of efficient and EF allocations have the same number
of components.

\subsection{Theorem 2}

\noindent Fix a single-valued, efficient rule $f$ meeting NE. Assume first $%
n=4$, $m=2$. Consider $\mathcal{Q}^{1}$ where, with the notation in the
previous proof, we have%
\begin{equation*}
r_{1}<r_{2}<\frac{1}{3}<1<3<r_{3}<r_{4}
\end{equation*}

By~(\ref{38}) and~(\ref{12}) $\mathcal{A}$ has three components: one
interior to $S^{1}$ (excluding the cut $z^{1/2}$), one around $z^{2/3}$
intersecting $S^{2}$ and $S^{3}$, and one interior to $S^{4}$ excluding $%
z^{3/4}$. Assume without loss that $f$ selects an allocation in the second
or third component just listed, and consider $\mathcal{Q}^{2}$ where $%
r_{1},r_{2}$ are unchanged but the new ratios $r_{3}^{\prime },r_{4}^{\prime
}$ are%
\begin{equation*}
r_{1}<r_{2}<3<r_{3}^{\prime }<1<r_{4}^{\prime }<\frac{1}{3}
\end{equation*}%
Here, again by~(\ref{38}) and~(\ref{12}), $\mathcal{A}$ has a single
component interior to $S^{1}$, the same as in $\mathcal{Q}^{1}$: none of the
cuts $z^{i/i+1}$ is in $\mathcal{A}$ anymore, and there is no component
interior to another $S^{i}$. When we decrease continuously $r_{3},r_{4}$ to $%
r_{3}^{\prime },r_{4}^{\prime }$, the allocation $z^{1/2}$ remains outside $%
\mathcal{A}$ and the component interior to $S^{1}$ does not move. Therefore
the allocation selected by $f$ cannot vary continuously in the ratios $r_{i}$%
, or in the underlying disutility matrix $u$.

We can clearly construct a similar pair of problems to prove the statement
when $n\geq 5$ and $m=2$. And for the case $m\geq 3$ we use the cloning
technique in Step 3 of the previous proof.

\end{document}